\documentclass[floatfix,twocolumn,showpacs,preprintnumbers,amsmath,amssymb,pra,superscriptaddress,longbibliography]{revtex4-1}
\usepackage{color}
\usepackage[usenames,dvipsnames,svgnames,table]{xcolor}
\usepackage[colorlinks=true,linkcolor=blue,urlcolor=blue,citecolor=blue]{hyperref}
\usepackage{mathtools}
\usepackage{graphicx}
\usepackage{dcolumn}
\usepackage{array}
\usepackage{lipsum}
\usepackage{bm}
\usepackage{subfigure}
\usepackage{amssymb}
\usepackage{multirow}
\usepackage{tabularx}
\usepackage{amsmath}
\usepackage{braket}
\usepackage{csquotes}
\graphicspath{{plots/}}
 \usepackage{lipsum}
\usepackage{mathrsfs}
\usepackage{MnSymbol}

\newcommand{\beq}{\begin{equation}}
\newcommand{\eeq}{\end{equation}}
\newcommand{\bea}{\begin{eqnarray}}
\newcommand{\eea}{\end{eqnarray}}

\begin{document}
\title{Fermionic physics from \emph{ab initio} path integral Monte Carlo simulations of fictitious identical particles}

\author{Tobias Dornheim}
\email{t.dornheim@hzdr.de}

\affiliation{Center for Advanced Systems Understanding (CASUS), D-02826 G\"orlitz, Germany}
\affiliation{Helmholtz-Zentrum Dresden-Rossendorf (HZDR), D-01328 Dresden, Germany}

\author{Panagiotis Tolias}
\affiliation{Space and Plasma Physics, Royal Institute of Technology (KTH), Stockholm, SE-100 44, Sweden}

\author{Simon Groth}

\affiliation{ Christian-Albrechts-Universit\"at zu Kiel,
 D-24098 Kiel, Germany}

\author{Zhandos~A.~Moldabekov}

\affiliation{Center for Advanced Systems Understanding (CASUS), D-02826 G\"orlitz, Germany}
\affiliation{Helmholtz-Zentrum Dresden-Rossendorf (HZDR), D-01328 Dresden, Germany}

\author{Jan Vorberger}

\affiliation{Helmholtz-Zentrum Dresden-Rossendorf (HZDR), D-01328 Dresden, Germany}

\author{Barak Hirshberg}

\affiliation{School of Chemistry, Tel Aviv University, Tel Aviv 6997801, Israel}
\affiliation{The Center for Computational Molecular and Materials Science, Tel Aviv University, Tel
Aviv 6997801, Israel}

\begin{abstract}
The \emph{ab initio} path integral Monte Carlo (PIMC) method is one of the most successful methods in statistical physics, quantum chemistry and related fields, but its application to quantum degenerate Fermi systems is severely hampered by an exponential computational bottleneck: the notorious fermion sign problem. Very recently, Xiong and Xiong [J.~Chem.~Phys.~\textbf{157}, 094112 (2022)] have suggested to partially circumvent the sign problem by carrying out simulations of fictitious systems guided by and interpolating continuous variable $\xi\in[-1,1]$, with the physical Fermi- and Bose-statistics corresponding to $\xi=-1$ and $\xi=1$. It has been proposed that information about the fermionic limit might be obtained by calculations within the bosonic sector $\xi>0$ combined with an extrapolation throughout the fermionic sector $\xi<0$, essentially bypassing the sign problem. Here, we show how the inclusion of the artificial parameter $\xi$ can be interpreted as an effective penalty on the formation of permutation cycles in the PIMC simulation. We demonstrate that the proposed extrapolation method breaks down for moderate to high quantum degeneracy. Instead, the method constitutes a valuable tool for the description of large Fermi-systems of weak quantum degeneracy. This is demonstrated for electrons in a 2D harmonic trap and for the uniform electron gas (UEG), where we find excellent agreement ($\sim0.5\%$) with exact configuration PIMC results in the high-density regime while attaining a speed-up exceeding eleven orders of magnitude. Finally, we extend the idea beyond the energy and analyze the radial density distribution (2D trap), as well as the static structure factor and imaginary-time density--density correlation function (UEG).
\end{abstract}

\maketitle

\section{Introduction\label{sec:introduction}}

Having been introduced in the 1960s for the description of ultracold $^4$He~\cite{Fosdick_PR_1966,Jordan_PR_1968}, the \emph{ab initio} path integral Monte Carlo (PIMC) method~\cite{Berne_JCP_1982,Takahashi_Imada_PIMC_1984,cep} has emerged as one of the most successful methods in statistical physics, quantum chemistry, and related disciplines. In principle, it allows for an exact solution of the full quantum $N$-body problem of interest without any empirical input such as the exchange--correlation functional required in density functional theory (DFT)~\cite{Jones_RMP_2015}. Furthermore, PIMC constitutes one of the few available tools that is capable to treat the interplay of quantum degeneracy effects, strong particle interactions and thermal excitations. Consequently, the PIMC method has been pivotal for our understanding of effects such as Bose-Einstein condensation~\cite{PhysRevB.72.014533,doi:10.7566/JPSJ.85.053001,PhysRevA.70.053614}, superfluidity~\cite{Pollock_PRB_1987,cep,Kwon_PRB_2006,Dornheim_PRB_2015}, and exotic supersolid behaviour~\cite{Saccani_Supersolid_PRL_2012}. Moreover, modern Monte Carlo sampling methods~\cite{boninsegni1,boninsegni2,Dornheim_PRB_nk_2021} allow for the quasi-exact simulation of up to $N\sim10^4$ particles.

Unfortunately, the situation is considerably more complicated in the case of Fermi statistics, where the antisymmetry with respect to the exchange of particle coordinates leads to a cancellation of positive and negative terms in the course of the evaluation of thermodynamic expectation values. This is the root cause of the notorious fermion sign problem (FSP)~\cite{troyer,dornheim_sign_problem,Dornheim_grand_2021} that leads to an exponential increase in the required compute time with increasing system size $N$ or decreasing temperature $T$.

This situation is highly unsatisfactory as quantum degenerate Fermi systems offer a wealth of interesting phenomena including the formation of Wigner molecules~\cite{Egger_PRL_1999,Reusch,PhysRevB.83.085409} and crystallization~\cite{Filinov_PRL_2001,PhysRevB.69.085116} at low density,
and the emergence of pairing-induced superfluidity and superconductivity~\cite{Takada_PRB_1993,Haussmann_PRA_2007,STRINATI20181,CHEN20051} at low temperatures. In addition, a particularly active field of research concerns the study of matter under extreme densities and temperatures~\cite{fortov_review,drake2018high,Dornheim_review}, as it occurs for instance in astrophysical objects such as giant planet interiors~\cite{Benuzzi_Mounaix_2014} and in inertial confinement fusion experiments e.g.~at the National Ignition Facility~\cite{Moses_NIF,hu_ICF}. This \emph{warm dense matter} (WDM) regime is defined by the absence of any small characteristic parameters, which rules out both weak-coupling expansions and ground-state methods~\cite{wdm_book,new_POP}.

The pressing need for accurate simulations of quantum degenerate Fermi systems at finite temperature has sparked a number of interesting developments in the field of thermal quantum Monte Carlo (QMC) methodologies over the last decades. Already thirty years ago, Ceperley~\cite{Ceperley1991} has suggested to extend the well-known \emph{fixed-node approximation}~\cite{anderson2007quantum,Anderson_fixe_nodes} to the finite-temperature PIMC method, which has allowed for simulations of a great variety of systems, including ultracold $^3$He~\cite{Ceperley_PRL_1992}, electrons in quantum dots~\cite{PhysRevB.96.205445} and a gamut of WDM systems~\cite{Militzer_PRE_2001,Militzer_2008,Militzer_Pollock_PRL_2002,Brown_PRL_2013}. While formally avoiding the FSP, this restricted PIMC method is based on an a-priori assumption about the nodal structure of the thermal density matrix, making it an uncontrolled approximation in practice. In fact, Schoof \emph{et al.}~\cite{Schoof_PRL_2015} have found systematic errors of up to $\sim10\%$ in the exchange--correlation energy in the high-density regime.

An interesting alternative is given by the \emph{blocking} paradigm, which is based on the alleviation of the FSP by combining a number of positive and negative terms into a single contribution. Notable examples for this idea are given by the multilevel blocking approach pioneered by Egger and co-workers~\cite{Egger_PRL_1998,Egger_PRL_1999,Egger_PRE_2000}, and the utilization of anti-symmetric imaginary-time propagators that has been explored by different groups~\cite{Chin_PRE_2015,Dornheim_NJP_2015,Dornheim_JCP_2015,Dornheim_CPP_2019,Filinov_CPP_2021,Filinov_PRE_2015}.

Other alternatives include methods formulated in the intrinsically anti-symmetric Fock space, such as the configuration PIMC method developed by Bonitz and co-workers~\cite{Schoof_CPP_2011,Schoof_CPP_2015,Schoof_PRL_2015,Groth_PRB_2016,groth_jcp,Yilmaz_JCP_2020}, and the density matrix QMC (DM-QMC) approach developed in the Foulkes group~\cite{Blunt_PRB_2014,Malone_JCP_2015,Malone_PRL_2016}. Both CPIMC and DM-QMC are efficient at high densities but become inefficient with increasing coupling strength. As a consequence, they are complementary to the direct PIMC method that is formulated in coordinate space, with the latter being efficient at strong coupling strength where quantum degeneracy effects are less important.
The combination of PIMC with CPIMC/DM-QMC thus allows one to cover substantial parts of the WDM parameter space~\cite{Dornheim_POP_2017,review}, which has resulted in a highly accurate parametrization of the exchange--correlation properties~\cite{groth_prl} of the uniform electron gas (UEG). These results have been confirmed very recently based on independent auxiliary field~\cite{Joonho_JCP_2021} and diagrammatic QMC simulations~\cite{Chen2019,Hou_PRB_2022}.

At the same time, it is important to note that both PIMC and CPIMC/DM-QMC calculations are still afflicted with the exponential scaling as a result of the FSP. To counteract the latter, it is often beneficial to carry out simulations in a modified configuration space where the sign problem is less severe (or, in the best case, entirely absent), followed by an extrapolation to the true fermionic limit. Successful implementations of this strategy include the \emph{kink extrapolation} in CPIMC~\cite{Groth_PRB_2016,Yilmaz_JCP_2020}, the \emph{initiator approximation} in DM-QMC~\cite{Malone_PRL_2016}, and the utilization of the Bogoliubov inequality and thermodynamic integration by Hirshberg and collaborators~\cite{Hirshberg_JCP_2020,Dornheim_JCP_2020}.  Very recently, Xiong and Xiong~\cite{Xiong_JCP_2022} have suggested to carry out path integral Molecular Dynamics (PIMD) simulations of particles whose canonical partition function features an external continuous variable $\xi\in[-1,1]$ (see Eq.~(\ref{eq:Z}) below) that interpolates between the physically meaningful cases of Fermi-statistics ($\xi=-1$) and Bose-statistics ($\xi=1$). More specifically, they have proposed to carry out simulations for $\xi\geq0$---which are not afflicted by any sign problem---and to subsequently extrapolate to the fermionic limit of $\xi=-1$ on the basis of an empirical quadratic expression. If successful, this approach would allow one to effectively circumvent the FSP, resulting in a de-facto exponential speed-up.

In this work, we demonstrate how the idea by Xiong and Xiong~\cite{Xiong_JCP_2022} can be understood as an effective exponential penalty on the formation of permutation cycles in the PIMC (or PIMD) simulation~\cite{Dornheim_permutation_cycles}, which constitute the root cause of the FSP. In practice, we demonstrate that the proposed extrapolation scheme works well for weakly quantum degenerate systems, but breaks down when the impact of fermionic anti-symmetry becomes substantial. This is demonstrated for two canonical test systems: 1) spin-polarized electrons in a 2D harmonic oscillator potential~\cite{Egger_PRL_1998,Egger_PRL_1999,Egger_PRE_2000,Dornheim_NJP_2015,Chin_PRE_2015,PhysRevB.96.205445,Xiong_JCP_2022,Dornheim_NJP_2022} and 2) the archetypal paramagnetic uniform electron gas model~\cite{loos,quantum_theory,review,moldabekov2023imposing}.
For the latter, we find excellent agreement to exact CPIMC reference data in the weakly degenerate high-density regime, while attaining a speed-up exceeding eleven orders of magnitude. 
Finally, we extend the considerations from the original Ref.~\cite{Xiong_JCP_2022} beyond the energy, and analyze the extrapolation scheme for the radial density profile of the 2D harmonic trap, as well as for the static structure factor and imaginary-time density-density correlation function of the UEG.

The paper is organized as follows: In Sec.~\ref{sec:theory}, we introduce the relevant theoretical background including a brief recap of the PIMC method and the sign problem resulting from its application to fermions. Sec.~\ref{sec:results} contains our extensive new PIMC simulation results for electrons in a 2D harmonic trap (\ref{sec:trap}) and for the warm dense UEG (\ref{sec:UEG}). The paper is concluded by a summary and discussion in Sec.~\ref{sec:summary}.

\section{Theory\label{sec:theory}}

Let us consider $N$ quantum particles (with $N=N^\uparrow+N^\downarrow$, and $N^\uparrow$ [$N^\downarrow$] denoting the number of spin-up [spin-down] particles) in a fixed volume $V$, and at a fixed inverse temperature $\beta=1/k_\textnormal{B}T$. The corresponding canonical partition function in the coordinate representation is given by 
\begin{widetext}
\begin{eqnarray}\label{eq:Z}
Z_{N,V,\beta} = \frac{1}{N^\uparrow!N^\downarrow!} \sum_{\sigma_{N^\uparrow}\in S_{N^\uparrow}}\sum_{\sigma_{N^\downarrow}\in S_{N^\downarrow}} \xi^{N_\textnormal{pp}} \int_V \textnormal{d}\mathbf{R}\ \bra{\mathbf{R}} e^{-\beta\hat{H}} \ket{\hat{\pi}_{\sigma_{N^\uparrow}}\hat{\pi}_{\sigma_{N^\downarrow}}\mathbf{R}}\ ,
\end{eqnarray}
\end{widetext}
with $\mathbf{R}=(\mathbf{r}_1,\dots,\mathbf{r}_N)^T$ containing the coordinates of all $N$ particles, $\sigma_{N^i}$ ($i\in\{\uparrow,\downarrow\}$) a particular element of the permutation group $S_{N^i}$, $\hat{\pi}_{\sigma_{N^i}}$ the corresponding permutation operator~\cite{Dornheim_permutation_cycles} and $N_{\mathrm{pp}}$ the number of pair permutations. In addition, the factor $\xi^{N_\textnormal{pp}}$ takes into account the impact of quantum statistics, with $\xi=1$ corresponding to bosons, $\xi=-1$ to fermions, and $\xi=0$ to hypothetical distinguishable quantum particles that are often being referred to as \emph{boltzmannons} in the literature.
Since a detailed derivation of the PIMC method has been presented elsewhere~\cite{cep}, we here restrict ourselves to a presentation of the final result of the partition function in the path integral representation. Specifically, after making use of the exact semi-group property of the density operator $\hat{\rho}=e^{-\beta\hat{H}}$ and the generalized Trotter formula~\cite{Trotter}, we get
\begin{widetext}
\begin{eqnarray}\label{eq:Z_final}
Z_{N,V,\beta} &=& \sumint_V \textnormal{d}\mathbf{X}\ \xi^{N_\textnormal{pp}} \bra{\mathbf{R}_0} e^{-\epsilon\hat{V}}e^{-\epsilon\hat{K}} \ket{\mathbf{R}_1}\bra{\mathbf{R}_1}\dots \ket{\mathbf{R}_{P-1}}\bra{\mathbf{R}_{P-1}}e^{-\epsilon\hat{V}}e^{-\epsilon\hat{K}}\ket{\hat{\pi}_{\sigma_{N^\uparrow}}\hat{\pi}_{\sigma_{N^\downarrow}}\mathbf{R}_0}\ \\ &=& \sumint_V \textnormal{d}\mathbf{X}\ W(\mathbf{X})\ , 
\end{eqnarray}
\end{widetext}
where the symbolic notation $\sumint_V \textnormal{d}\mathbf{X}$, with $\mathbf{X}=(\mathbf{R}_0,\dots,\mathbf{R}_{P-1})^T$, also includes the summation over all possible permutations of particle coordinates. Moreover, we have split the full Hamiltonian into a potential and kinetic part, $\hat{H} = \hat{V} + \hat{K}$, and introduced the reduced inverse temperature $\epsilon=\beta/P$. We note that the introduced primitive factorization $e^{-\epsilon\hat{H}}\approx e^{-\epsilon\hat{V}}e^{-\epsilon\hat{K}}$ becomes exact in the limit of large $P$ as $\mathcal{O}\left(P^{-2}\right)$~\cite{kleinert2009path}. Higher-order factorizations have been discussed in the literature~\cite{sakkos_JCP_2009}, but their application is most efficient at low temperatures, where the FSP generally prevents PIMC simulations of Fermi systems due to the associated exponential bottleneck.

From a physical perspective, Eq.~(\ref{eq:Z_final}) can be interpreted as follows: Each quantum particle is now represented by a set of $P$ particle coordinates that are diffusing throughout the imaginary time~\cite{Dornheim_insight_2022,Dornheim_PTR_2022}, with $\Delta \tau=-i\hbar\epsilon$ the so-called imaginary-time step. The basic idea of the PIMC method is to stochastically sample all possible paths according to their corresponding configuration weight $P(\mathbf{X}) = W(\mathbf{X})/Z_{N,V,\beta}$, which can be done efficiently using different implementations of the celebrated Metropolis algorithm~\cite{metropolis}.

The method of Xiong and Xiong~\cite{Xiong_JCP_2022}, that will be presented in detail in Section~\ref{xiong}, employed PIMD for sampling the fermionic partition function. Therefore, we briefly outline key features of fermionic and bosonic PIMD and refer the readers to the references below for the full details.
A straightforward implementation of PIMD simulations of bosonic particles is not feasible, since the sum over all permutations in Eq.~(\ref{eq:Z_final}) leads to an exponential scaling with the number of particles. However, Hirshberg et al.~recently showed that the potential and forces required for performing PIMD simulations of bosonic systems can be evaluated recursively, reducing the computational cost from exponential to cubic~\cite{Hirshberg_PNAS_2019,Woo_PRL_2022}. 
Feldman and Hirshberg then reduced the computational cost further~\cite{feldman2023quadratic}, from cubic to quadratic, allowing the first simulations of thousands of particles using PIMD at a similar cost to PIMC.
PIMD is numerically exact, like PIMC, but offers a very different way of sampling both configuration and permutation space. 

The recursive PIMD algorithm was also used to simulate fermions through a reweighting procedure (see section~\ref{sign}), as long as the FSP is not too severe~\cite{Hirshberg_JCP_2020}, and molecular dynamics-based enhanced sampling techniques were employed to alleviate it for small systems~\cite{Dornheim_JCP_2020}. Throughout the paper, we present PIMC simulations results but the conclusions are valid for both PIMD and PIMC.

\subsection{PIMC and the fermion sign problem}
\label{sign}

Within the fermionic sector $\xi\in[-1,0)$, in general, and at the fermionic limit of $\xi=-1$, in particular, the configuration weight $P(\mathbf{X})$ can attain both positive and negative values, which rules out the straightforward interpretation as a probability distribution. Instead, we have to generate the configurations $\mathbf{X}$ according to the modified weight function $W'(\mathbf{X})=|W(\mathbf{X})|$ that is normalized by the corresponding modified partition function
\begin{eqnarray}
    Z'_{N,V,\beta} = \sumint_V \textnormal{d}\mathbf{X}\ W'(\mathbf{X})\ .
\end{eqnarray}
Following this route, the expectation value of a given observable $\hat{O}$ is then given by
\begin{eqnarray}\label{eq:ratio}
    \braket{\hat{O}} = \frac{\braket{\hat{O}\hat{S}}'}{\braket{\hat{S}}'}\ ,
\end{eqnarray}
with $\braket{\dots}'$ denoting the evaluation of the expectation value in the modified ensemble, and with the denominator in Eq.~(\ref{eq:ratio}) being known as the \emph{average sign},
\begin{eqnarray}\label{eq:sign}
    S &=& \braket{\hat{S}}' \nonumber\\
    &=&\frac{1}{Z'_{N,V,\beta}} \sumint_V \textnormal{d}\mathbf{X}\  \frac{W(\mathbf{X})}{|W(\mathbf{X})|}|W(\mathbf{X})| \nonumber\\ 
    &=& \frac{Z_{N,V,\beta}}{Z'_{N,V,\beta}}\ .
\end{eqnarray}
In practice, Eq.~(\ref{eq:sign}) constitutes a direct measure for the degree of cancellation between positive and negative terms in the PIMC simulation. Indeed, the associated relative statistical uncertainty of the expectation value Eq.~(\ref{eq:ratio}) is inversely proportional to the sign,
\begin{eqnarray}\label{eq:error_with_sign}
    \frac{\Delta O}{O} \sim \frac{1}{S\sqrt{N_\textnormal{MC}}} \ ,
\end{eqnarray}
which, due to the central limit theorem, can only be compensated by increasing the number of Monte Carlo samples as $1/\sqrt{N_\textnormal{MC}}$.
The well-known exponential decrease of $S$ with parameters such as the system size $N$ or the inverse temperature $\beta$ then directly leads to the exponential computational bottleneck, i.e., the FSP~\cite{troyer,dornheim_sign_problem}.

From a practical perspective, it is easy to see that the modified configuration space defined by $Z'_{N,V,\beta}$ for $\xi<0$ directly corresponds to the analogous partition function for $|\xi|$. Therefore, we get results both for $|\xi|$ and $-|\xi|$ from the same PIMC simulation.

\subsection{Artificial interpolating variable and extrapolation procedure}\label{xiong}

Let us next discuss the role of the artificial interpolating variable $\xi$. In the fermionic sector $\xi\in[-1,0)$, we carry out a bosonic PIMC simulation and subsequently reconstruct the fermionic expectation value of interest via Eq.~(\ref{eq:ratio}); the latter is subject to the full FSP, leading to a potentially vanishing signal-to-noise ratio for large systems and low temperatures. For $\xi=0$, the configuration weight always vanishes except for those configurations with zero pair permutations, $N_\textnormal{pp}=0$. This set-up thus corresponds to a PIMC simulation of distinguishable boltzmannons. In the bosonic sector $\xi\in(0,1]$, pair permutations are allowed in the PIMC simulation with a non-zero weight, but the $\xi^{N_\textnormal{pp}}$ factor servers as an additional, artificial weight function that exponentially suppresses the formation of permutation cycles. Considering the full range of $\xi\in[-1,1]$ thus gives one a continuous transition between the three physically motivated cases of $\xi=-1,0,1$.

Since PIMC simulations for $\xi\geq0$ are not afflicted with any sign problem, Xiong and Xiong~\cite{Xiong_JCP_2022} have suggested to extrapolate from the range of $\xi\in[0,1]$ to the fermionic limit of $\xi=-1$, using the empirical quadratic polynomial form~\cite{Xiong_JCP_2022}
\begin{eqnarray}\label{eq:fit}
    O(\xi) = a_O + b_O\xi + c_O\xi^2 \ .
\end{eqnarray}
As concluded in the following Sec.~\ref{sec:results}, the simple functional form of Eq.~(\ref{eq:fit}) works surprisingly well for weak degrees of quantum degeneracy, but it breaks down when the impact of quantum statistics becomes substantial.

It is beyond the scope of present investigation to establish a connection between this simulation-motivated interpolation procedure and existing physics-motivated fractional exclusion statistics. Nevertheless, similarities are apparent and possible synergies could be exploitable in more refined interpolation techniques and extrapolating procedures.

\section{Results\label{sec:results}}

All PIMC results that are presented in this work have been obtained using the extended ensemble approach introduced in Ref.~\cite{Dornheim_PRB_nk_2021}, which is a canonical adaption of the worm algorithm by Boninsegni \emph{et al.}~\cite{boninsegni1,boninsegni2}. We use $P=100-200$ high-temperature factors within the primitive factorization, and the convergence with $P$ has been carefully checked.

\subsection{Electrons in a 2D harmonic oscillator\label{sec:trap}}

As the first system, we consider $N$ spin-polarized, i.e., $N=N^\uparrow$ and $N^\downarrow=0$, electrons in two-dimensional harmonic confinement. Note that we assume oscillator units throughout this section. The corresponding Hamiltonian is given by
\begin{eqnarray}\label{eq:Hamiltonian_HO}
\hat{H} = -\frac{1}{2}\sum_{l=1}^N \nabla^2_l + \frac{1}{2} \sum_{l=1}^N \hat{\mathbf{r}}_l^2 + \sum^N_{l<k} \frac{\lambda}{|\hat{\mathbf{r}}_l-\hat{\mathbf{r}}_k|}\ 
\end{eqnarray}
with $\lambda\geq0$ being the dimensionless coupling parameter. From a physical perspective, Eq.~(\ref{eq:Hamiltonian_HO}) is often used as a simple model for electrons in quantum dots~\cite{Reimann_RMP_2002} and offers a plethora of potentially interesting effects, including an abnormal \emph{negative} superfluid fraction~\cite{Dornheim_NJP_2022}, as well as the formation of Wigner molecules~\cite{Egger_PRL_1999} and crystallization~\cite{Filinov_PRL_2001} at low temperatures. In addition, we mention the manifestation of the \emph{quantum breathing mode} that has been investigated in detail by Abraham and co-workers~\cite{PhysRevLett.111.256801,Abraham_CPP_2014}. 
Moreover, Eq.~(\ref{eq:Hamiltonian_HO}) constitutes a widely used benchmark for the application and development of novel methodologies~\cite{Dornheim_NJP_2015,Chin_PRE_2015,PhysRevB.96.205445,Xiong_JCP_2022,Egger_PRL_1998,Egger_PRL_1999,Egger_PRB_2005,Hirshberg_JCP_2020,Dornheim_JCP_2020}.

\begin{figure}
\includegraphics[width=0.462\textwidth]{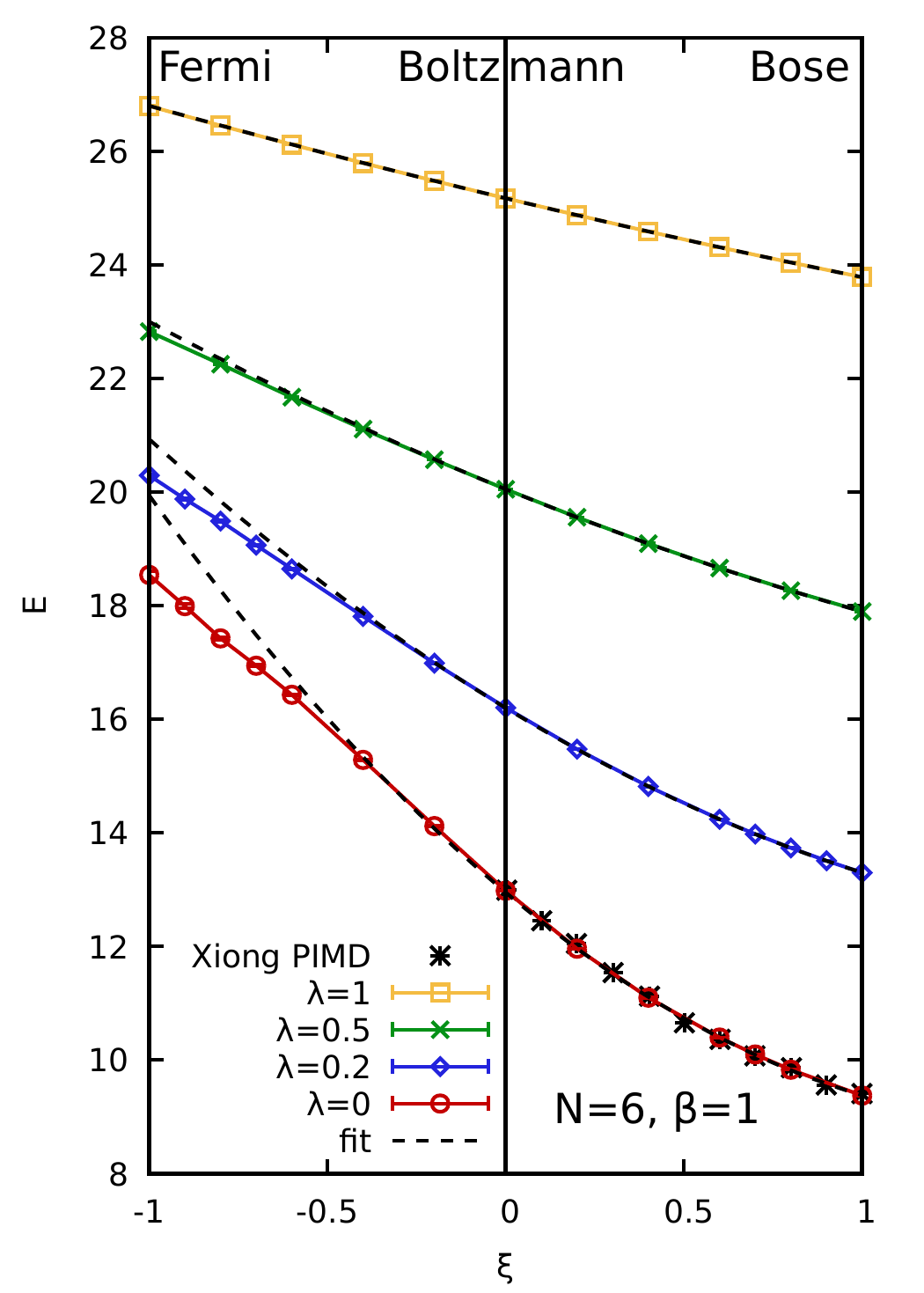}
\caption{\label{fig:lambda}
The PIMC results for the total energy $E$ of $N=6$ spin-polarized electrons in a 2D harmonic trap at $\beta=1$ and for different values of the coupling parameter $\lambda$. The dashed black lines depict fits to the sign-problem free domain $\xi\in[0,1]$ via Eq.~(\ref{eq:fit}). The black stars depict PIMD data taken from Xiong and Xiong~\cite{Xiong_JCP_2022}.
}
\end{figure} 

\begin{figure*}
\includegraphics[width=0.462\textwidth]{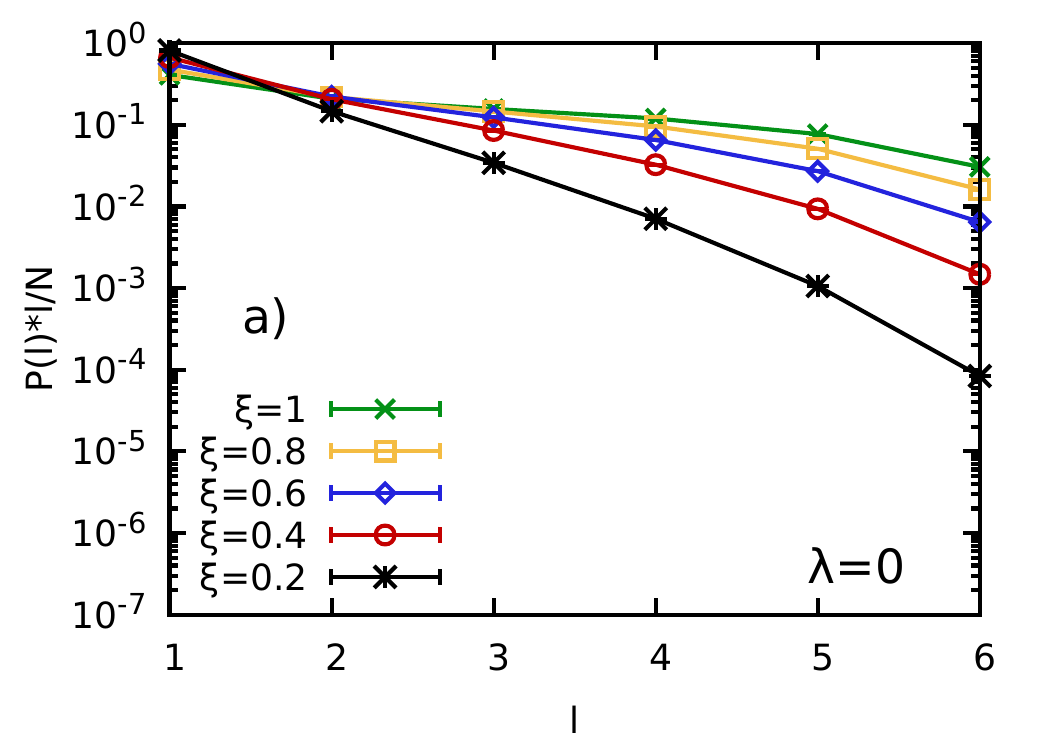}\includegraphics[width=0.462\textwidth]{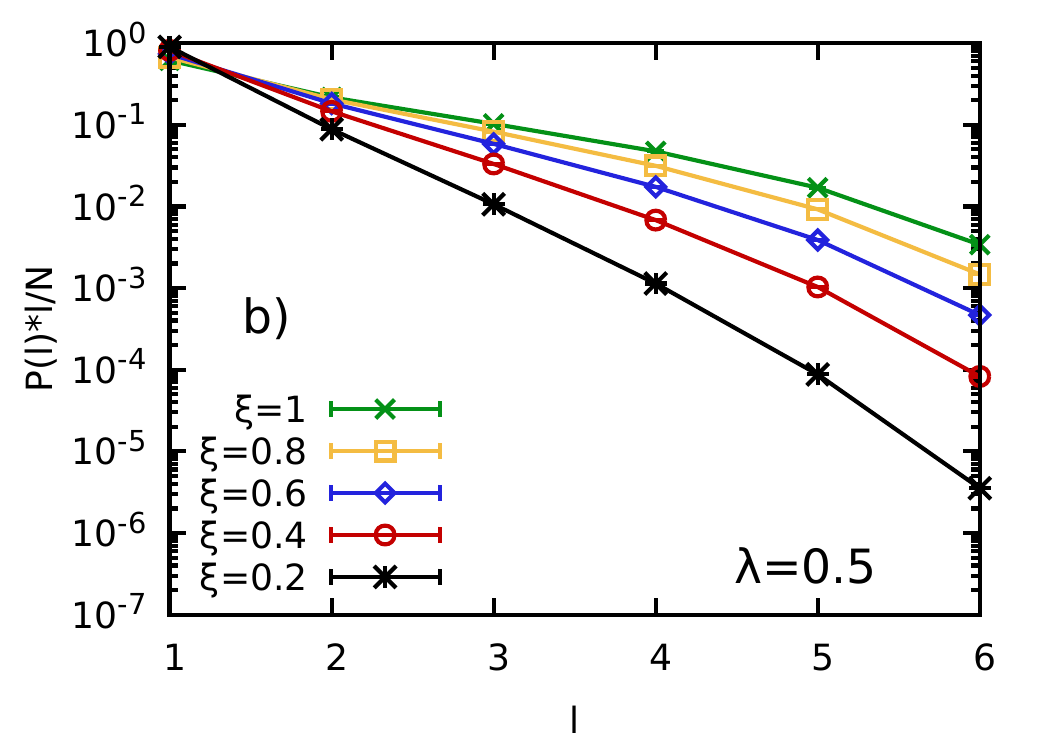}\\\includegraphics[width=0.462\textwidth]{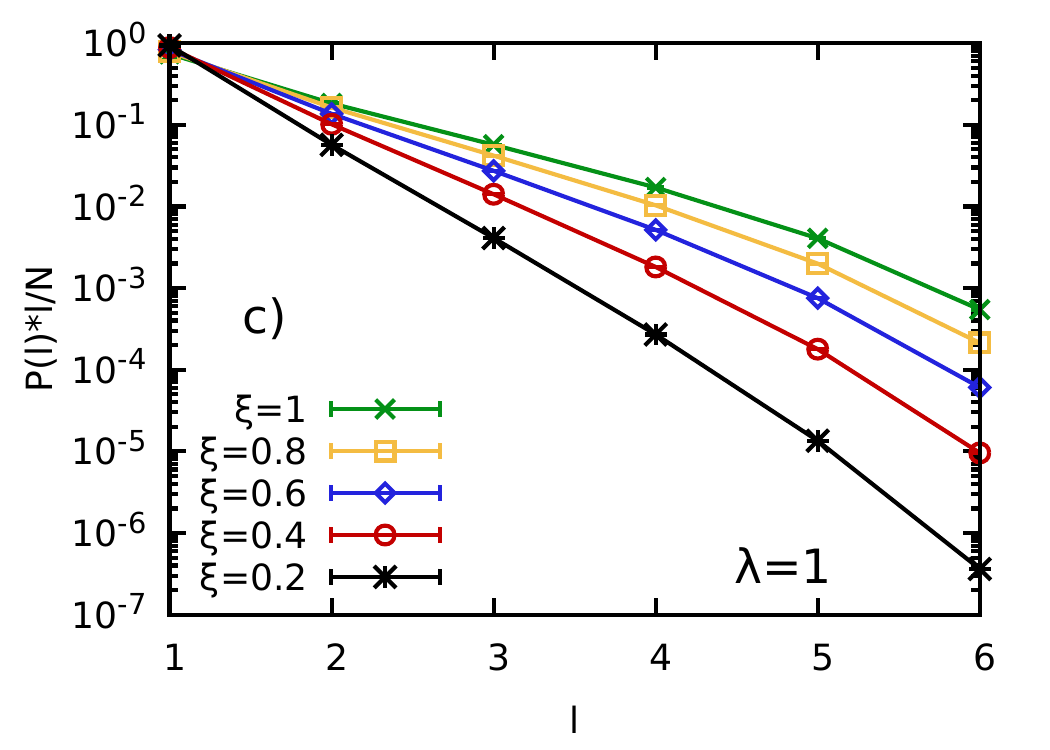}\includegraphics[width=0.462\textwidth]{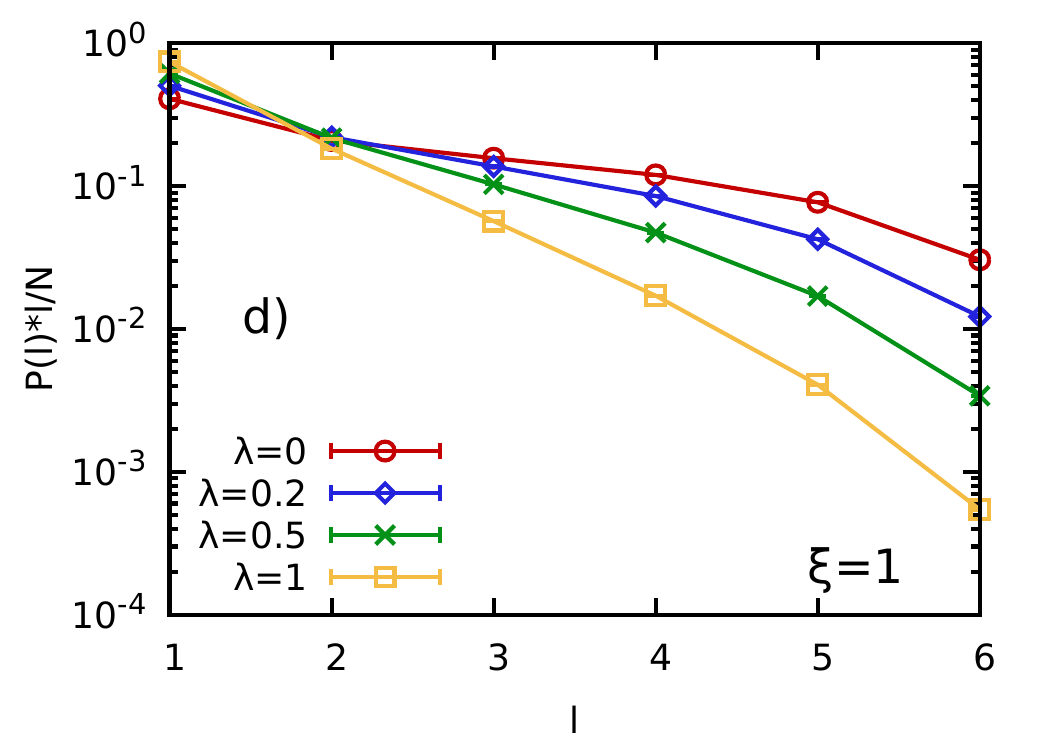}
\caption{\label{fig:Xiong_Fig6_permutations}
Distribution of permutation cycles in PIMC simulations of $N=6$ spin-polarized electrons in a 2D harmonic trap with $\beta=1$.
Panels a)-c) compare results for different $\xi$ for $\lambda=0,0.5,1$, and panel d) compares results for different $\lambda$ in the limit of $\xi=1$.
}
\end{figure*} 

\begin{figure}
\includegraphics[width=0.462\textwidth]{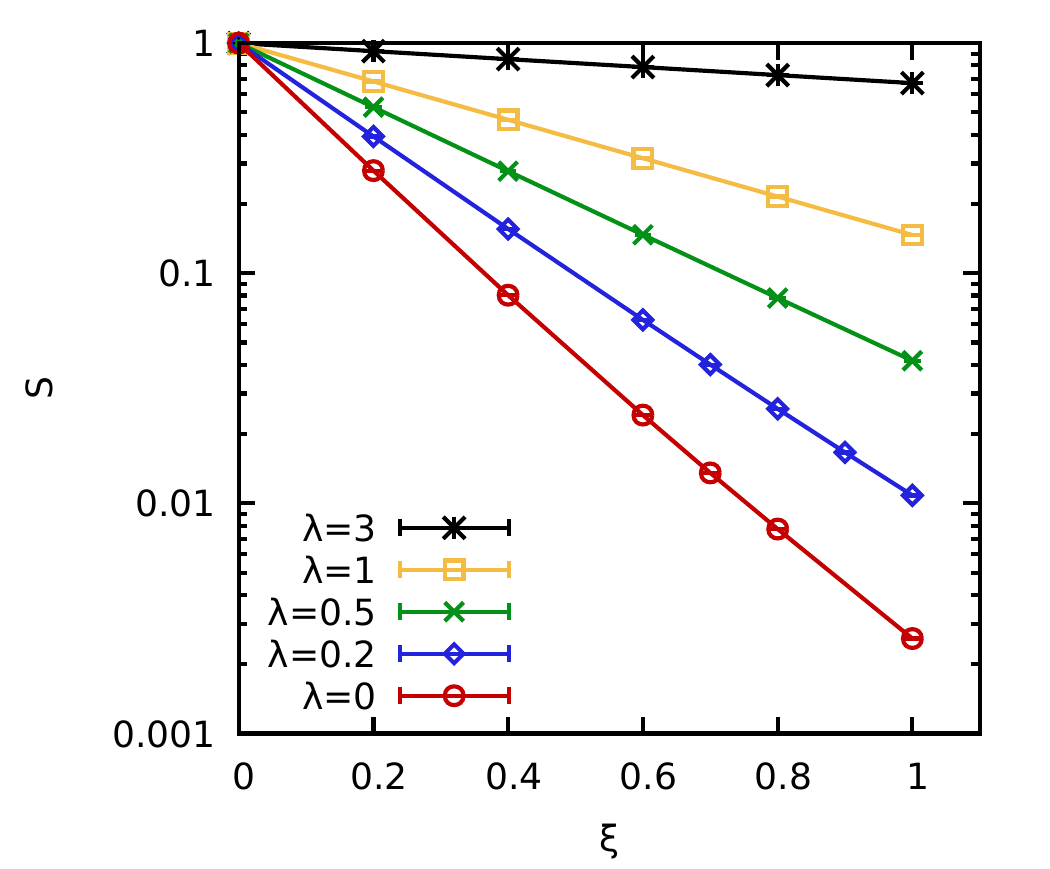}
\caption{\label{fig:sign_lambda}
The average sign of the PIMC simulations depicted in Fig.~\ref{fig:lambda}.
}
\end{figure} 

In Fig.~\ref{fig:lambda}, new PIMC results are shown for the total energy $E$ as a function of the continuous interpolating variable $\xi$ for $N=6$, $\beta=1$, and different $\lambda$ values. Naturally, the total energy in the system increases with $\lambda$, as the particles push each other away from the center of the trap compared to the ideal case of $\lambda=0$. For $\lambda=1$, the impact of quantum statistics is already significant, but the difference between the fermionic ($\xi=-1$) and the bosonic limits ($\xi=1$) does not exceed $10\%$. 
The corresponding dashed black line has been fitted to the yellow squares in the sign-problem free domain of $\xi\in[0,1]$ via the quadratic Eq.~(\ref{eq:fit}), and predicts the full range of $\xi<0$ with remarkable accuracy. This finding empirically substantiates the functional form of Eq.~(\ref{eq:fit}), and supports the idea proposed in Ref.~\cite{Xiong_JCP_2022}.

Reduction of the interaction strength by a factor of two leads to the green crosses. In this case, the fit to the $\xi\in[0,1]$ domain still nicely reproduces its input data, but there appears to be a noticeable inaccuracy in the fermionic limit of $\xi=-1$. This trend is further exacerbated for $\lambda=0.2$ (blue diamonds), and even more so for $\lambda=0$ (red circles), where we find an extrapolation error of $\sim10\%$ in the fermionic limit. For completeness, we include the PIMD results for $\lambda=0$ from Ref.~\cite{Xiong_JCP_2022} as the black stars; they are in perfect agreement with our PIMC results, cross-validating our respective implementations. 

To understand the apparent breakdown of the extrapolation governed by Eq.~(\ref{eq:fit}), let us first inspect the functional form of our PIMC results over the entire depicted $\xi$-range.
Clearly, there appears an inflection point around $\xi\simeq-0.5$ that cannot be reproduced by a quadratic polynomial dependency.
Hence, the extrapolation tends to systematically overestimate the effect of quantum statistics when the degree of quantum degeneracy is high.

Aiming to shed light to the mechanism behind the manifestation of the quantum degeneracy effects, we analyze the distribution of permutation cycles of length $l$ in Fig.~\ref{fig:Xiong_Fig6_permutations}. Specifically, we plot the probability of a particle to be involved in a permutation cycle of length $l$, $P(l)l/N$, such that the sum over all $l$ is normalized to unity. Panel a) corresponds to the ideal case of $\lambda=0$, with the green crosses depicting results for $\xi=1$. Note that $P(l)$ is an observable of the modified configuration space defined by $Z'_{N,V,\beta}$ and, thus, is the same for $|\xi|$ and $-|\xi|$. While the probability to find a single particle that is not involved in any exchange is comparably larger compared to $l>1$, the distribution is quite flat. This indicates proximity to the low-temperature regime, where permutation cycles of all lengths are equally likely~\cite{krauth2006statistical}. Upon decreasing $|\xi|$, the distribution of permutation cycles becomes increasingly less flat as pair permutations are exponentially penalized in the PIMC simulations by the additional factor of $\xi^{N_\textnormal{pp}}$.
Heuristically, it then makes sense that the nontrivial interplay of permutation cycles of different lengths indicated by the flat distribution of $P(l)l/N$ at $|\xi|=1$ cannot be inferred from the boltzmannon-to-boson $\xi$-domain. The large impact of quantum statistics at these parameters can also directly be seen in Fig.~\ref{fig:sign_lambda}, where we show the average sign $S$ [cf.~Eq.~(\ref{eq:sign})] of our PIMC simulations. We find an exponential decrease of $S$ with $\xi$ that is most pronounced for $\lambda=0$, as it is expected. For larger $\lambda$, the electrons get effectively separated by the Coulomb repulsion, which leads to a reduced probability to form permutation cycles in PIMC~\cite{Dornheim_permutation_cycles}.

Let us next consider Fig.~\ref{fig:Xiong_Fig6_permutations}b) and c), where we repeat our analysis of permutation cycle distributions for $\lambda=0.5$ and $\lambda=1$. In both cases, we find that $P(l)l/N$ is substantially less flat in the limit of $|\xi|=1$ compared to $\lambda=0$, which explains 1) the less pronounced impact 
of quantum statistics in these cases and 2) the better performance of the extrapolation.
Finally, we show results for different values of $\lambda$ (at $\xi=1$) in panel d), which further substantiates our previous conclusions.

\begin{figure}
\includegraphics[width=0.462\textwidth]{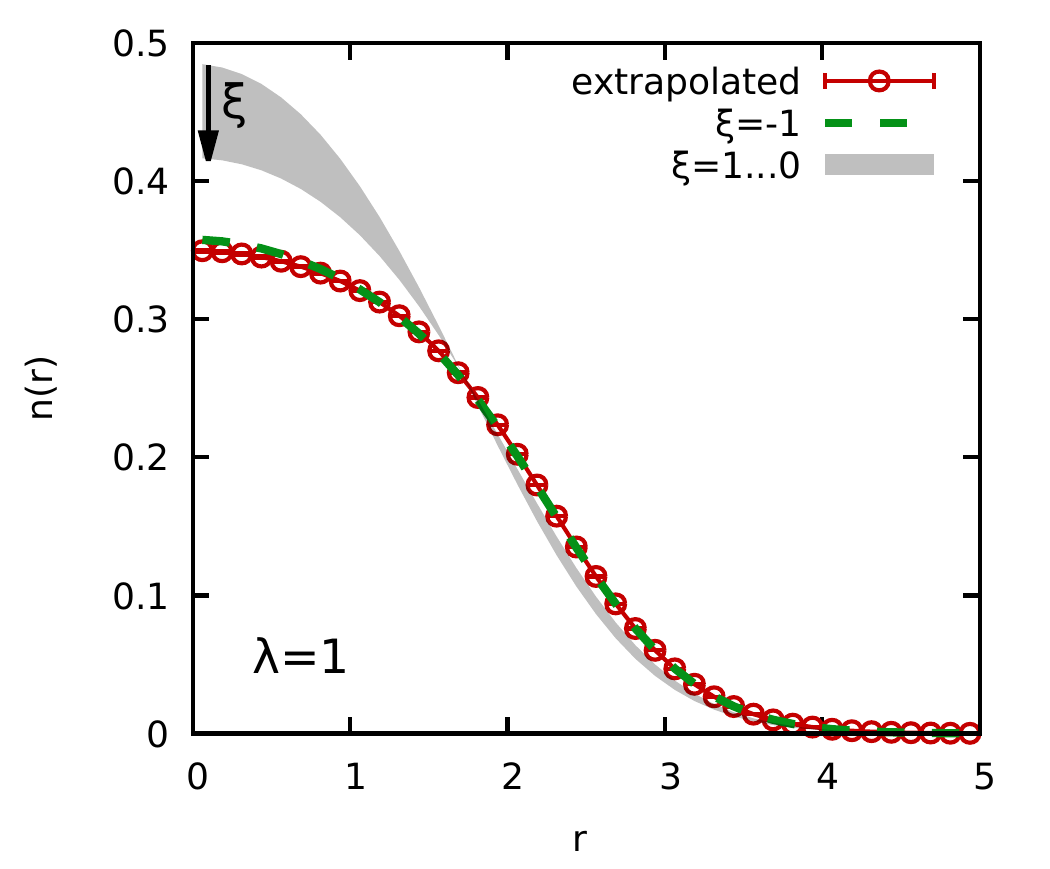}\\\includegraphics[width=0.462\textwidth]{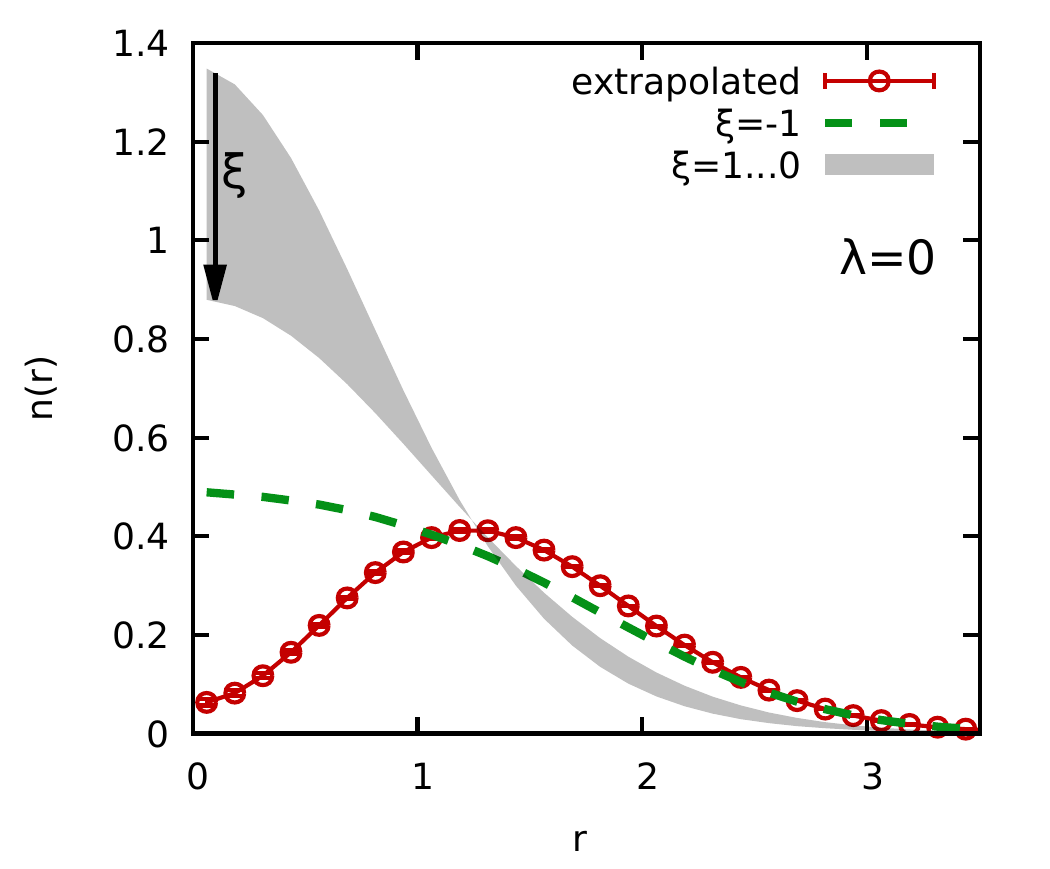}
\caption{\label{fig:Extrapolate_density}
Application of the $\xi$-extrapolation idea to the radial density distribution $n(r)$ for $N=6$ and $\beta=1$ with the top and bottom panels corresponding to $\lambda=1$ and $\lambda=0$.
}
\end{figure} 

An additional interesting question is given by the applicability of the extrapolation scheme to different observables beyond the total energy $E$. In Fig.~\ref{fig:Extrapolate_density}a) we consider the radial density distribution $n(r)$ for $N=6$, $\beta=1$, and $\lambda=1$, i.e., in a regime where the extrapolation is expected to work well. The shaded grey area encloses PIMC results in the sign-problem free domain of $\xi\in[0,1]$, with the black arrow indicating the gradient from the bosonic to the boltzmannonic case.  In addition, the dashed green line depicts our fermionic PIMC results for $\xi=-1$. Under these conditions, the FSP is not too severe and we can converge the PIMC results to obtain a numerically-exact result. From the point of physical intuition, these results can be understood as follows. When compared to distinguishable quantum particles, bosons tend to cluster more closely around the center of the harmonic trap, whereas fermions are repelled away from the center by the effective degeneracy pressure~\cite{Dornheim_CPP_2019}. The red circles correspond to the extrapolated results [using the parabolic form given in Eq.~(\ref{eq:fit})] that have been obtained exclusively based on input data for $\xi\in[0,1]$ via Eq.~(\ref{eq:fit}). First and foremost, we find very good qualitative agreement between the red and green curves over the entire $r$-range. In particular, the extrapolation gives virtually exact results for $r\gtrsim 1$; small inaccuracies only appear around the center of the trap where the degree of local quantum degeneracy is expected to be the highest.

In Fig.~\ref{fig:Extrapolate_density}b), we repeat the same analysis for the case of $\lambda=0$. Owing to the overall increased impact of quantum statistics, we observe larger differences between bosons, boltzmannons, and fermions. Note the average sign of $S\approx 2.5\times10^{-3}$ for $\xi=-1$, which makes the PIMC simulations involved, but still feasible. Let us conclude this exploration of the radial density distribution by considering the extrapolation from the sign-problem free domain to the fermionic limit depicted by the red circles. It is evident that the extrapolation substantially overestimates the impact of the degeneracy pressure, resulting in a spurious shell structure around $r\approx1.2$ and a minimum around the center of the trap; this trend is absent from the exact fermionic PIMC results. In contrast, the extrapolation works well for large $r\gtrsim2.5$, even though the impact of quantum statistics on the radial density distribution is of the order of $100\%$ in this regime.

\begin{figure}
\includegraphics[width=0.462\textwidth]{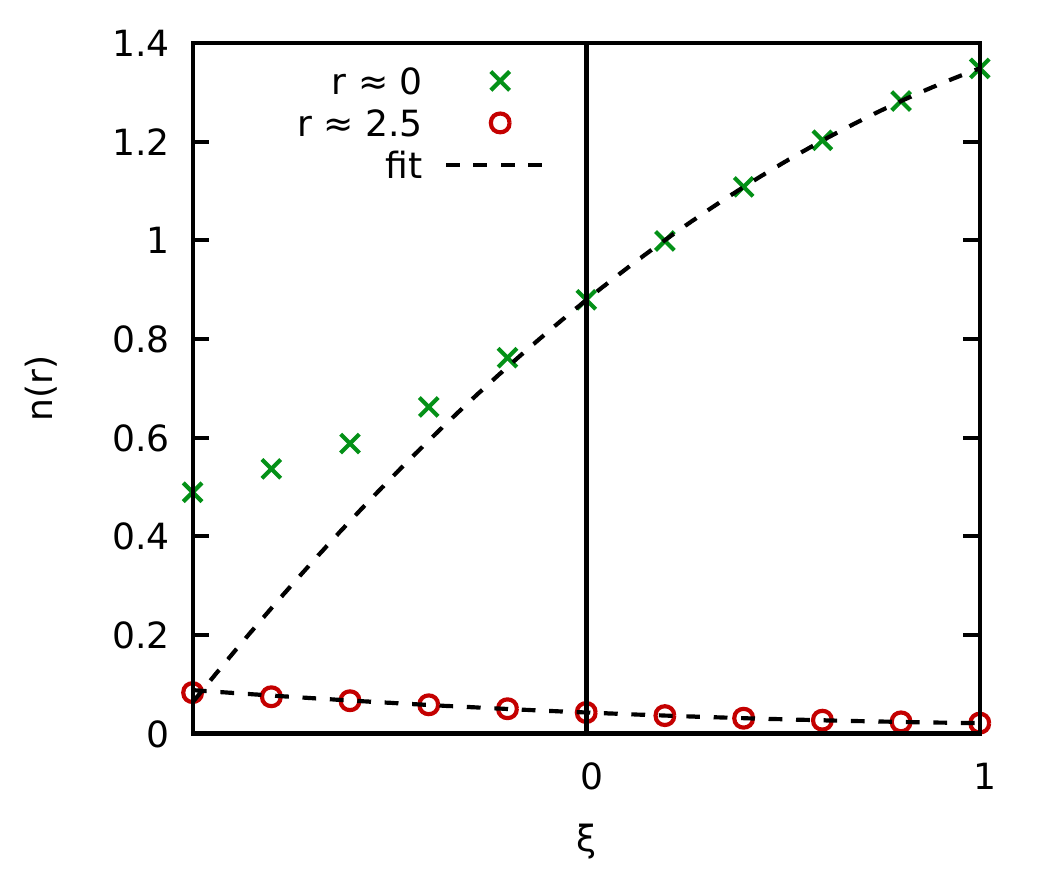}
\caption{\label{fig:Xiong_Fig6_iLine}
Extrapolation of the radial density distribution for $N=6$, $\beta=1$, and $\lambda=0$, cf.~Fig.~\ref{fig:Extrapolate_density}b), for two values of $r$.
}
\end{figure} 

This can be discerned particularly well in Fig.~\ref{fig:Xiong_Fig6_iLine}, where we depict the $\xi$-extrapolation for two representative values of $r$. Specifically, the green crosses correspond to $r\approx0$. Here, the actual PIMC results exhibit an inflection point that is not reproduced by the extrapolation, leading to the observed overestimation of quantum degeneracy effects. In contrast, the red circles corresponding to $r\approx2.5$ do not exhibit such a feature and the extrapolation from the sign-problem free domain to the fermionic limit works well.

\begin{figure}
\includegraphics[width=0.462\textwidth]{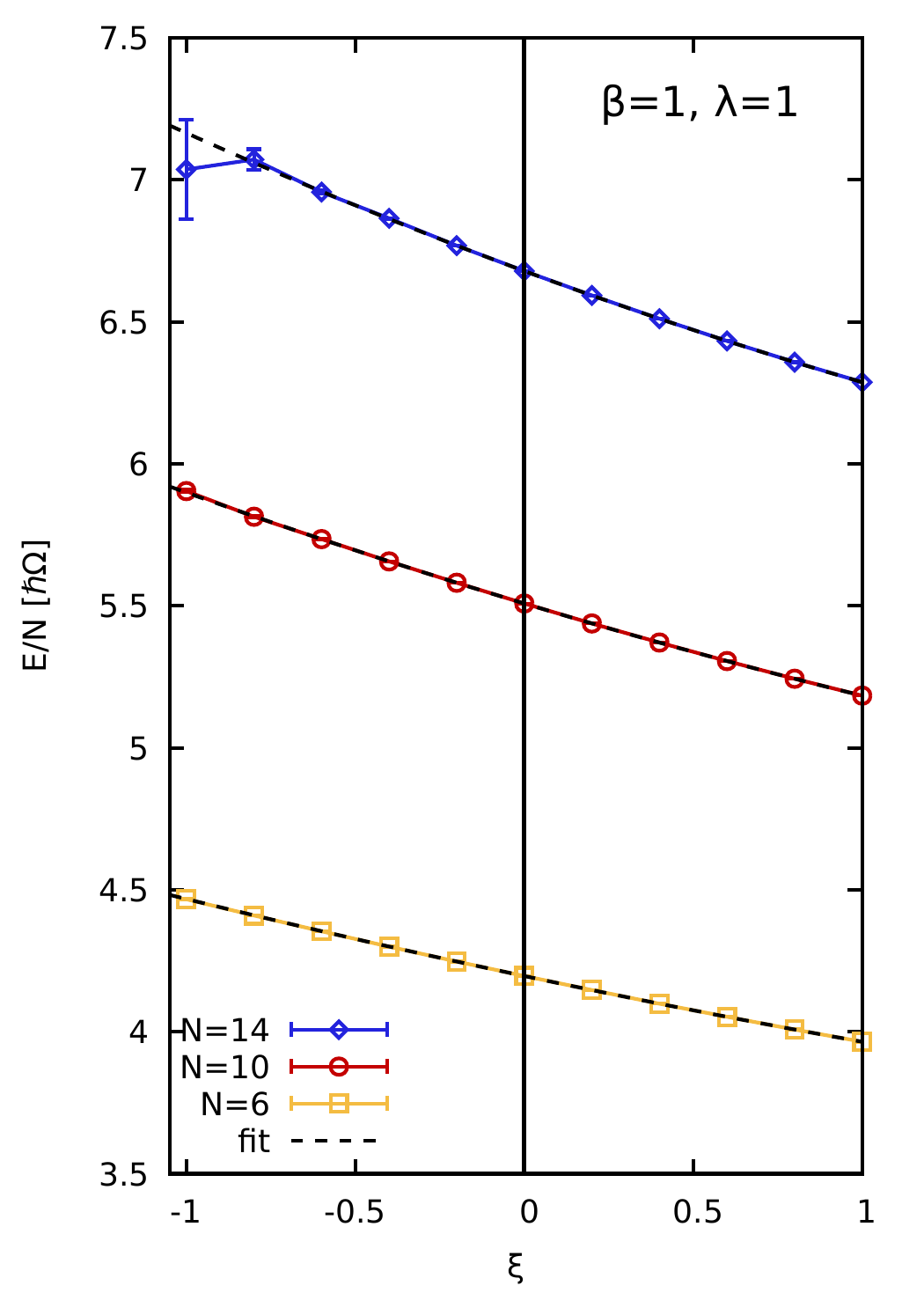}
\caption{\label{fig:N}
Total energy per particle for $N=6$ (yellow squares), $N=10$ (red circles), and $N=14$ (blue diamonds) spin-polarized electrons in a 2D harmonic trap with $\beta=1$ and $\lambda=1$. The dashed black curves have been obtained based on fits in the sign-problem free domain $\xi\in[0,1]$ via Eq.~(\ref{eq:fit}).
}
\end{figure} 

\begin{figure}
\includegraphics[width=0.462\textwidth]{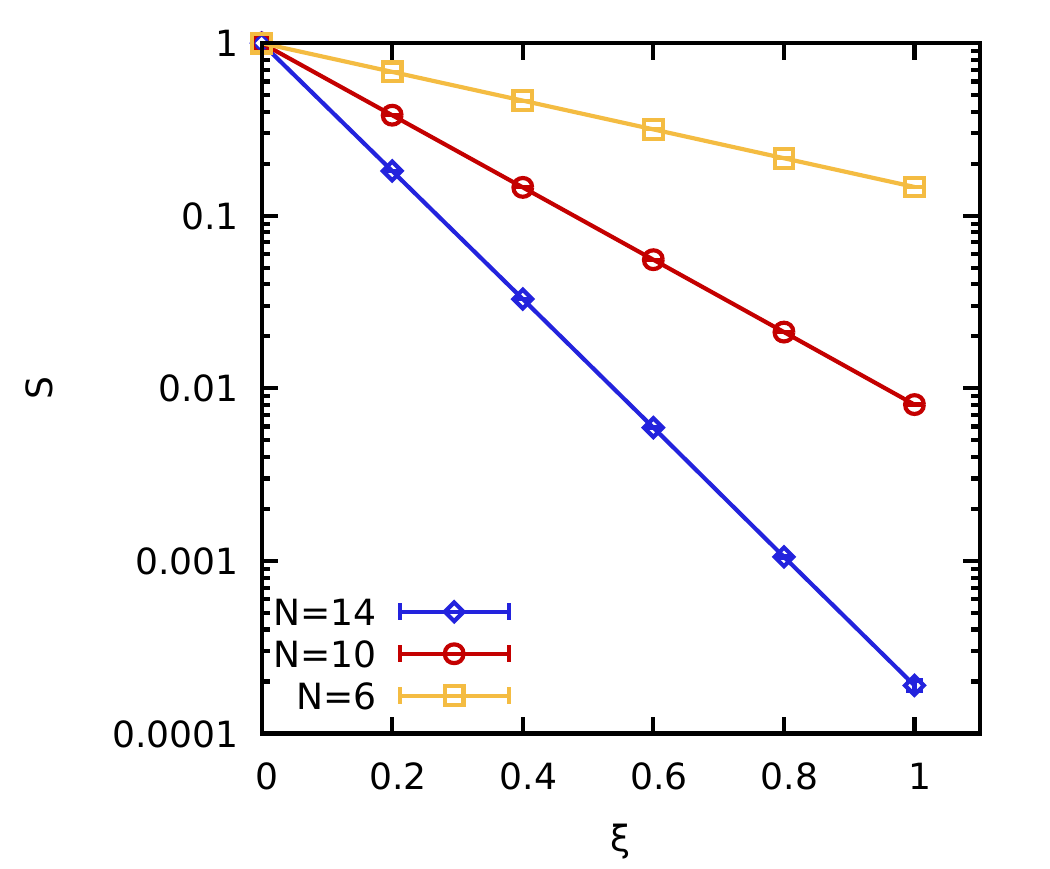}
\caption{\label{fig:sign_lambda1_N}
Average sign for the PIMC simulations presented in Fig.~\ref{fig:N}.
}
\end{figure}

Let us conclude our investigation of electrons in the harmonic oscillator potential with an effort to utilize the conclusions drawn so far to make fermionic PIMC simulations more efficient.  In particular, it is reasonable to assume that the extrapolation is reliable in situations with weak quantum degeneracy that are still subject to the exponential computational bottleneck due to the scaling with the system size $N$. If true, the proposed extrapolation scheme might lead to an exponential increase in efficiency in this regime.
To test this hypothesis, we show results for the total energy per particle for $N=6$ (yellow squares), $N=10$ (red circles), and $N=14$ (blue diamonds) electrons at $\beta=1$ and $\lambda=1$ in Fig.~\ref{fig:N}. As before, the dashed black lines correspond to fits governed by the quadratic Eq.~(\ref{eq:fit}) and constructed exclusively from the sign-problem free domain, $\xi\in[0,1]$. 
Indeed, we find that the extrapolation works extremely well in all three considered cases. For $N=6$ and $N=10$, accurate PIMC simulations are feasible over the entire relevant $\xi$-range and 
we find perfect agreement between the fits and the simulation results even in the fermionic limit of $\xi=-1$.
For $N=14$, the sign problem is substantially more severe, and we find an average sign of $S\sim10^{-4}$ for $\xi=-1$, cf.~Fig.~\ref{fig:sign_lambda1_N}. Nevertheless, the fit reproduces all PIMC results within the given level of statistical uncertainty, and the extrapolated result at $\xi=-1$ seems to be reliable.

\subsection{Uniform electron gas\label{sec:UEG}}

\begin{figure}
\includegraphics[width=0.462\textwidth]{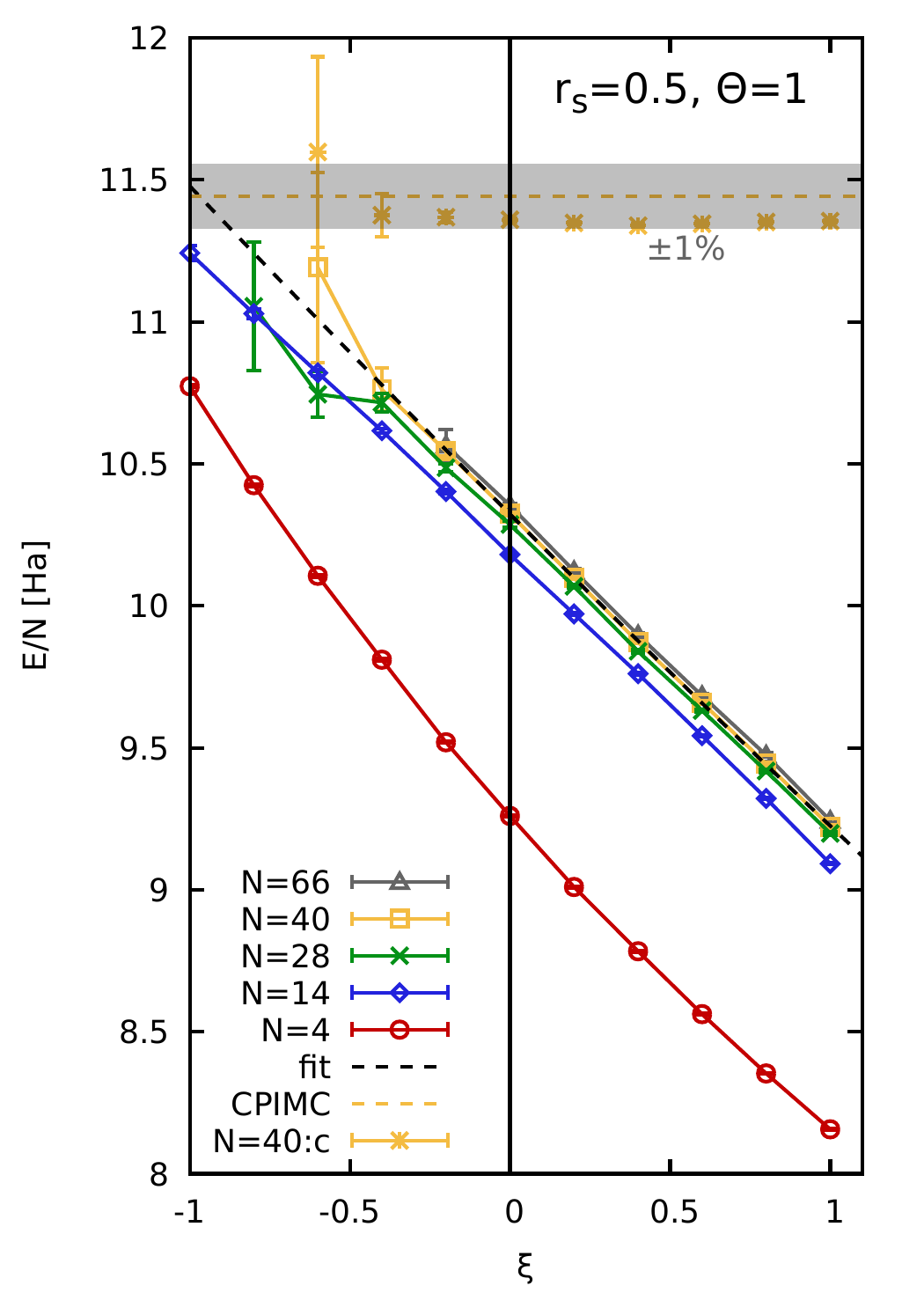}
\caption{\label{fig:UEG_PerParticle}
Energy per particle of the spin-unpolarized UEG at the state $r_s=0.5$, $\Theta=1$. The grey triangles, yellow squares, green crosses, blue diamonds, and red circles illustrate direct PIMC results for $N=66$, $N=40$, $N=28$, $N=14$, and $N=4$, respectively. The dashed black line corresponds to a fit to the $N=40$ data in the FSP free domain of $\xi\in[0,1]$ via the quadratic Eq.~(\ref{eq:fit}). The horizontal dashed yellow line corresponds to the exact CPIMC value in the limit of $\xi=-1$. The yellow stars have been obtained by adding a quantum-statistics correction based on $N=14$ simulations to the yellow squares, see main text. The shaded grey area indicates deviations of $\pm1\%$ from the CPIMC results and has been included as a reference.
}
\end{figure} 

The uniform electron gas is the second model system that is considered in this study. Having originally been introduced as the simplest realistic description of the conduction electrons in metals~\cite{mahan1990many}, the UEG has emerged as one of the most important model systems in statistical physics, quantum chemistry and related fields~\cite{loos,quantum_theory,review}. Indeed, it has only been the accurate parametrization of various properties of the UEG~\cite{Perdew_Zunger_PRB_1981,Perdew_Wang_PRB_1992,vwn,farid,cdop,Gori-Giorgi_PRB_2000,groth_prl,ksdt,status,dornheim_ML,Dornheim_PRB_2021} based on quantum Monte Carlo simulations in the ground state~\cite{Ceperley_Alder_PRL_1980,moroni,moroni2,Spink_PRB_2013} and at finite temperature~\cite{Brown_PRL_2013,Dornheim_JCP_2015,Groth_PRB_2016,Dornheim_PRB_2016,Malone_JCP_2015,Malone_PRL_2016,dornheim_prl} that has facilitated the success of density functional theory regarding the description of real materials~\cite{Jones_RMP_2015}. Of particular importance are the linear~\cite{moroni,moroni2,bowen2,Chen2019,Kukkonen_PRB_2021,dornheim_ML,dornheim_HEDP,dornheim_electron_liquid,Dornheim_PRL_2020_ESA,Dornheim_PRR_2022} and nonlinear~\cite{Dornheim_PRL_2020,Dornheim_JCP_ITCF_2021,Dornheim_JPSJ_2021,jctc_22,Dornheim_PRR_2021,Tolias_EPL_2023} density responses to external perturbations, as well as the dynamic structure factor~\cite{dornheim_dynamic,dynamic_folgepaper,Dornheim_Nature_2022,Takada_PRB_2016,Hamann_PRB_2020} that is central to the interpretation of X-ray Thomson scattering experiments~\cite{siegfried_review,kraus_xrts,Dornheim_review}.

From a practical perspective, the UEG is conveniently characterized by two dimensionless parameters, both of the order of one in the WDM regime~\cite{Ott2018}: 1) The density parameter $r_s=d/a_\textnormal{B}$ corresponds to the Wigner-Seitz radius expressed in units of the Bohr radius, and plays the role of the quantum coupling parameter. Hence, the UEG attains behaves as an ideal Fermi gas for $r_s\to0$, then it becomes a strongly coupled electron liquid for $r_s\sim10$~\cite{dornheim_electron_liquid,Tolias_JCP_2021,Tolias_JCP_2023} and eventually changes phase into a Wigner crystal for $r_s\gtrsim100$~\cite{PhysRevB.69.085116,Azadi_Wigner_2022}. 2) The reduced temperature $\Theta=k_\textnormal{B}T/E_\textnormal{F}$, with $E_\textnormal{F}$ the usual Fermi energy~\cite{quantum_theory}, measures the degree of quantum degeneracy, with $\Theta\ll1$ and $\Theta\gg1$ corresponding to the fully degenerate and semi-classical~\cite{Dornheim_HEDP_2022} limits. 
For completeness, it is pointed out that a third relevant parameter is given by the degree of spin-polarization, but we restrict ourselves to a fully unpolarized system with $N^\uparrow=N^\downarrow$ in this work. All simulation results will be presented in Hartree atomic units throughout this section.

\begin{figure}
\includegraphics[width=0.462\textwidth]{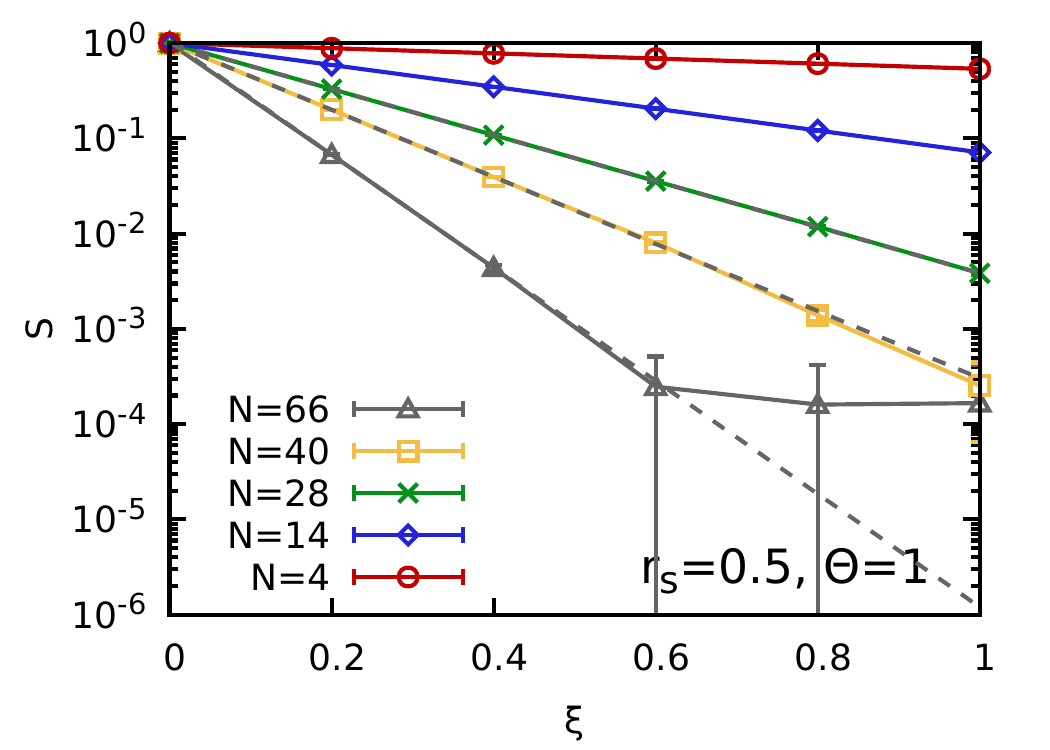}
\caption{\label{fig:UEG_sign_theta1}
Average sign of the PIMC simulations of the spin-unpolarized UEG at $r_s=0.5$, $\Theta=1$, cf.~Fig.~\ref{fig:UEG_PerParticle}.
}
\end{figure}

In Fig.~\ref{fig:UEG_PerParticle}, we show the energy per particle of the UEG in the high density regime with $r_s=0.5$ and $\Theta=1$. Such conditions are highly relevant for contemporary WDM research and can be realized for example in inertial confinement fusion experiments at the National Ignition Facility (NIF)~\cite{Moses_NIF}. In addition, the condition $\Theta=1$ implies partial degeneracy, which, based on our results for the harmonic oscillator discussed in the previous section, makes the proposed extrapolation method promising. For $N=4$ (red circles) and $N=14$ (blue diamonds), direct PIMC simulations are feasible over the entire $\xi$-range, and we find no inflection point in either case. For $N=28$ (green crosses), simulations become unfeasible around $\xi=0.6$, and this trend is exacerbated for $N=40$ (yellow squares) and $N=66$ (grey triangles), as it is expected. The corresponding values of the average sign $S$ are shown in Fig.~\ref{fig:UEG_sign_theta1}, with the sign vanishing within the given level of uncertainty for $N=66$ and $\xi\geq0.6$. Also note the exponential decrease of $S$ with $\xi$, in agreement with the trends observed for the 2D harmonic trap in Sec.~\ref{sec:trap}

Apparently, the conditions investigated in Fig.~\ref{fig:UEG_PerParticle} are well suited for the extrapolation method to provide accurate results that are beyond the capability of the direct PIMC method due to the FSP. To rigorously assess this notion, exact configuration PIMC results for $N=40$ have been included as a dashed horizontal line. Remarkably, there is a very good agreement with the extrapolated result, i.e., the $\xi=-1$ limit of the dashed black curve based on a quadratic Eq.~(\ref{eq:fit}) fit to the PIMC results in the FSP free domain. The attained accuracy is substantially better than $1\%$, with the latter being shown as the shaded grey area around the CPIMC result. It is noted that this merely serves as a guide-to-the-eye, since the statistical error of the CPIMC result is orders of magnitude smaller, making it an ideal benchmark for the present case.

An alternative extrapolation procedure can be gleamed from the close qualitative agreement of the $\xi$-dependence of the PIMC data sets shown in Fig.~\ref{fig:UEG_PerParticle} for $N\geq14$. It is thus tempting to assume that one can combine PIMC simulation results at $\xi\geq0$ for a large system with the $\xi$-dependence for a particle number for which PIMC calculations are possible for all $\xi$, $N=14$ being the obvious choice in the present case. This strategy also allows for a straightforward heuristic physical justification, as it constitutes an effective combination of a large-scale description of Coulomb correlation effects in the FSP free domain with a description of exchange effects obtained on a smaller length scale. This makes intuitive sense, since one would not expect long-range correlations to strongly depend on the particular quantum statistics, whereas, conversely, fermionic exchange effects are empirically known to be a rather short-range phenomenon~\cite{Kohn_PNAS_2005}. Implementation of this strategy for a particular system size $N$ and smaller reference system $N_\textnormal{ref}$ leads to the expression
\begin{eqnarray}\label{eq:corrected}
\left.\frac{E_N}{N}\right|_{\xi=-1} = \left.\frac{E_N}{N}\right|_{\xi} + \left(
\left.\frac{E_{N_\textnormal{ref}}}{{N_\textnormal{ref}}}\right|_{\xi=-1} - \left.\frac{E_{N_\textnormal{ref}}}{{N_\textnormal{ref}}}\right|_{\xi} 
\right)\ .
\end{eqnarray}
The results for $N=40$ and $N_\textnormal{ref}=14$ have been included as the yellow stars in Fig.~\ref{fig:UEG_PerParticle}. First and foremost, we find that the thus corrected results do not depend on $\xi$ within the given Monte Carlo error bars. At the same time, we also find significant deviations of $\sim0.8\%$ to the CPIMC reference result, which is somewhat worse than the direct extrapolation via Eq.~(\ref{eq:fit}). Interestingly, this implies that there is a small residual interplay of quantum statistical effects with medium-range correlations that is not fully covered by the simulation results for $N_\textnormal{ref}=14$.

\begin{figure}
\includegraphics[width=0.462\textwidth]{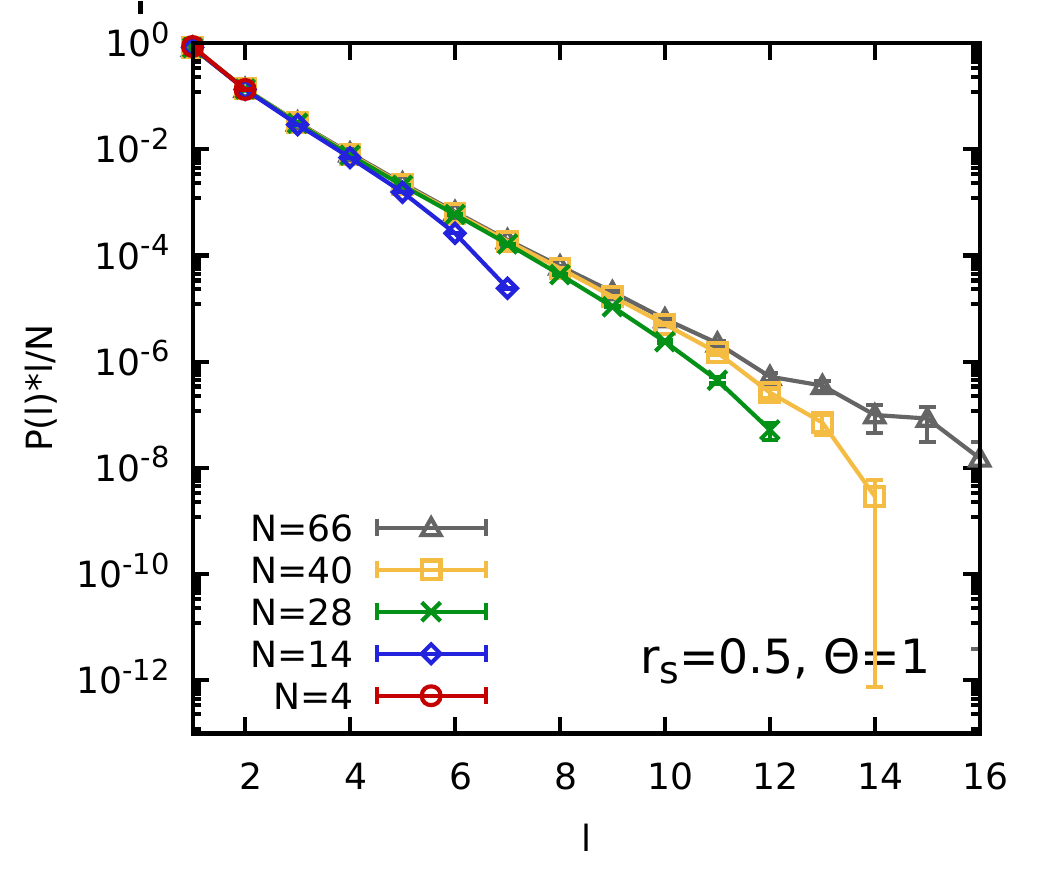}
\caption{\label{fig:UEG_permutations_theta1}
Distribution of permutation cycles (at $\xi=1$) for the spin-unpolarized UEG at $r_s=0.5$, $\Theta=1$, and for different electron numbers $N$.
}
\end{figure} 

To gain additional insights into the analysis presented in Fig.~\ref{fig:UEG_PerParticle}, we show the corresponding distributions of permutation cycles in Fig.~\ref{fig:UEG_permutations_theta1}. Note that the maximum permutation length is given by $N/2$ in the paramagnetic case as only electrons of the same spin-orientation can be involved into a common exchange cycle~\cite{Dornheim_permutation_cycles}. Clearly, there appears no plateau in $P(l)l/N$ as in the quantum degenerate case (i.e., $\lambda=0$) investigated in Fig.~\ref{fig:Xiong_Fig6_permutations} above, which further substantiates the applicability of the extrapolation method. Second, we find a very similar exponential decay for all investigated particle numbers, which indicates that the degree of quantum degeneracy is indeed nearly independent of the system size for the UEG, even though the FSP becomes exponentially more severe with increasing $N$ (Fig.\ref{fig:UEG_sign_theta1}). Finally, a $N$-dependent drop is observed in $P(l)l/N$ for $l\sim N/2$. This is a finite-size effect, likely connected to the small residual error of Eq.~(\ref{eq:corrected}).

\begin{figure}
\includegraphics[width=0.462\textwidth]{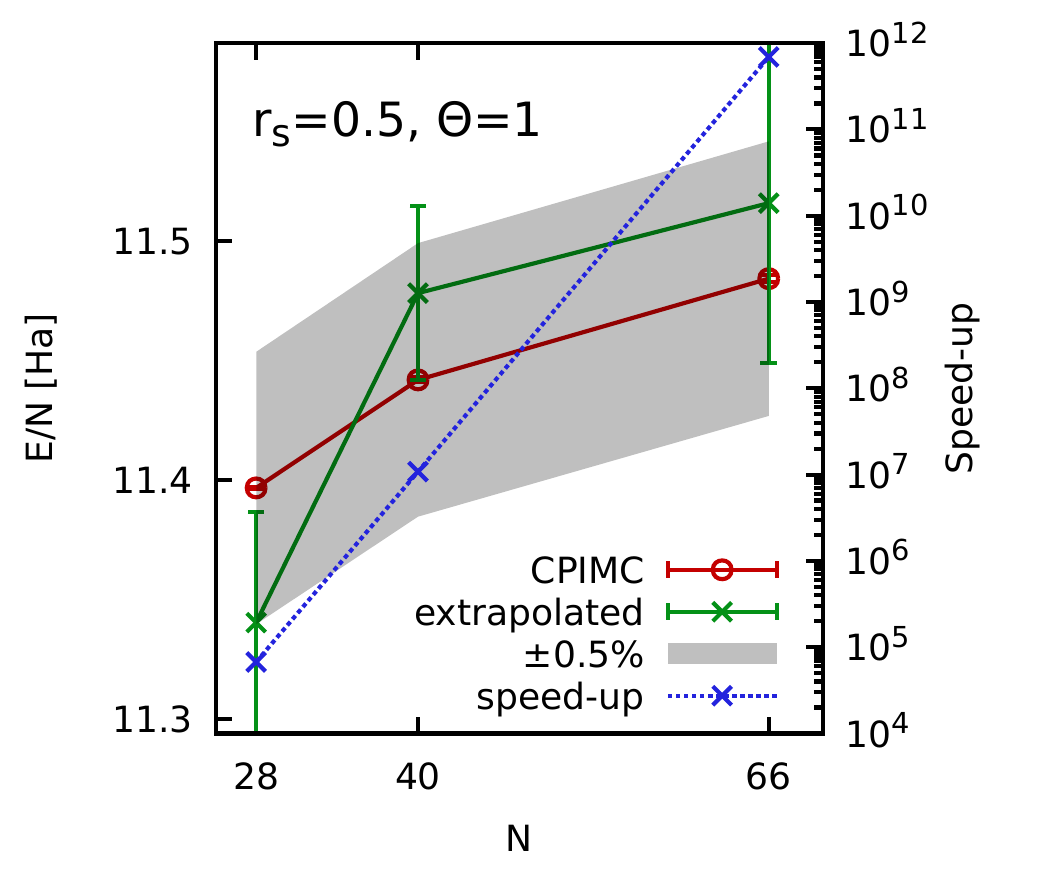}
\caption{\label{fig:Extrapolated_UEG_data}
Benchmarking of the extrapolation from the FSP free domain (green crosses) against exact CPIMC results (red circles) at $r_s=0.5$, $\Theta=1$, and for different electron numbers $N$. The shaded grey area indicates an interval of $\pm0.5\%$ around the CPIMC results and has been included as a guide-to-the-eye. The blue crosses indicate the respective speed-up (right $y$-axis) of the extrapolation method compared to full PIMC simulations that are subject to the FSP.
}
\end{figure} 

Having established the PIMC $\xi$-extrapolation as a reliable tool at these parameters, we provide a more quantitative analysis of its suitability to simulate large system sizes in Fig.~\ref{fig:Extrapolated_UEG_data}. The green crosses indicate extrapolated results that are based on fits restricted to the FSP free domain of $\xi\geq0$ via the quadratic Eq.~(\ref{eq:fit}) with the corresponding extrapolation uncertainty depicted in the form of error bars. These results are in excellent agreement with the exact CPIMC benchmark data for all $N$, with a typical deviation of $\lesssim0.5\%$, see the shaded grey area introduced as a guide to the eye. Finally, the blue crosses (right $y$-axis) depict the respective speed-up $c=1/S^2$ of the extrapolation method with respect to direct PIMC simulations that are subject to the full FSP; it directly follows from Eq.~(\ref{eq:error_with_sign}). Due to the exponential scaling of the FSP, the sign-problem free extrapolation attains an exponentially increasing speed-up that exceeds eleven orders of magnitude for $N=66$. Substantially larger particle numbers can be covered in future works, of possible high value as a further cross-check of existing finite-size correction schemes~\cite{Chiesa_PRL_2006,dornheim_prl,Dornheim_JCP_2021,Dornheim_PRE_2020,Holzmann_PRB_2016,groth_jcp}.

\begin{figure}
\includegraphics[width=0.462\textwidth]{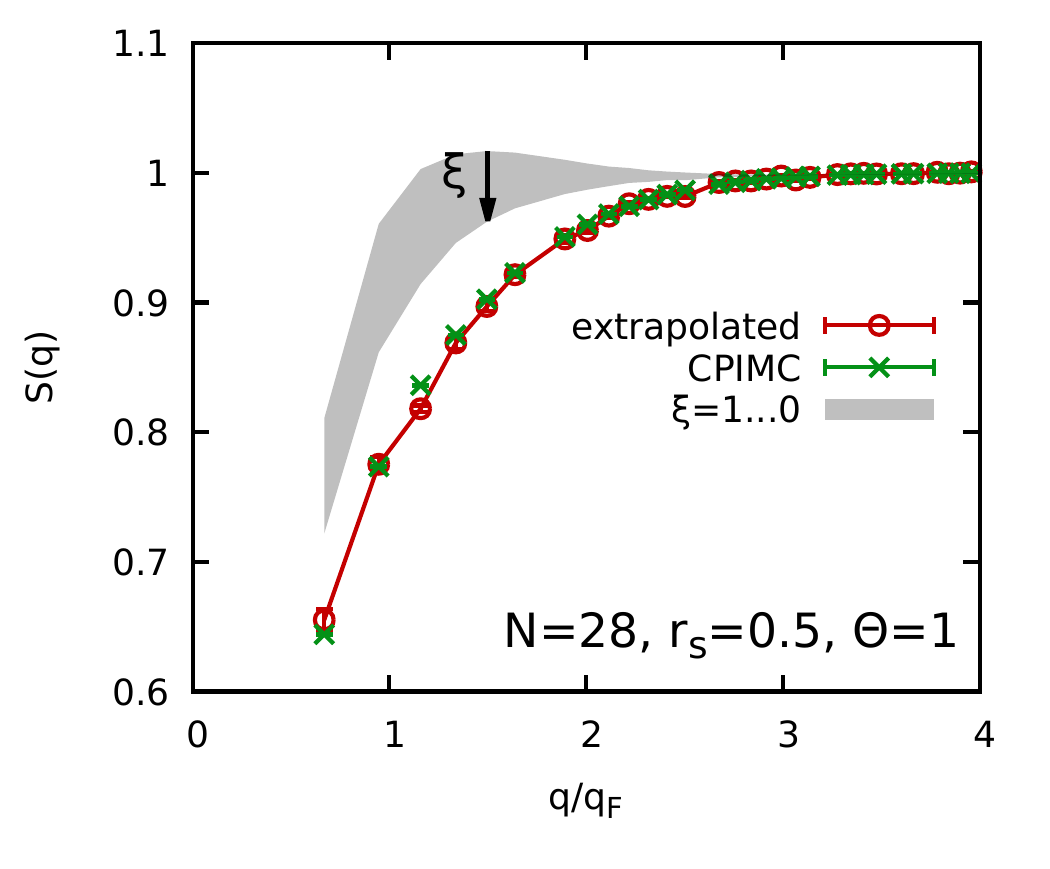}
\caption{\label{fig:UEG_SF_N28}
Static structure factor $S(q)$ of the spin-unpolarized UEG at $r_s=0.5$, $\Theta=1$ and for $N=28$. Shaded grey area: PIMC results for the FSP free domain of $\xi\geq0$; red circles: corresponding extrapolation from the grey area to the fermionic limit of $\xi=-1$ via the quadratic Eq.~(\ref{eq:fit}); green crosses: exact CPIMC reference data. 
}
\end{figure} 

Let us proceed with our investigation of the $r_s=0.5$ and $\Theta=1$ state by considering the extrapolation of the static structure factor $S(q)$ depicted in Fig.~\ref{fig:UEG_SF_N28}. For completeness, we note that $S(q)$ of the UEG only depends on the absolute value of the wave vector $q=|\mathbf{q}|$. In addition, the wave number axis has been rescaled with respect to the Fermi wave number $q_\textnormal{F}\sim1/r_s$~\cite{quantum_theory}, which provides an appropriate reference scale thereby allowing for comparisons between different values of the density parameter $r_s$. The shaded grey area encloses our PIMC results for the FSP free domain of $\xi\geq0$, with the solid black arrow indicating the gradient from the bosonic to the boltzmannonic regime. The red circles show the correspondingly extrapolated results based on the usual fits via the quadratic Eq.~(\ref{eq:fit}). They are in excellent agreement with the exact CPIMC reference data (green crosses) over the entire range of wave numbers. In practice, this means that the remarkable speed-up of the extrapolation method reported in Fig.~\ref{fig:Extrapolated_UEG_data} can even be extended to observables beyond the total energy, which makes the approach a flexible tool for the investigation of large, weakly quantum degenerate WDM systems.

\begin{figure}
\includegraphics[width=0.462\textwidth]{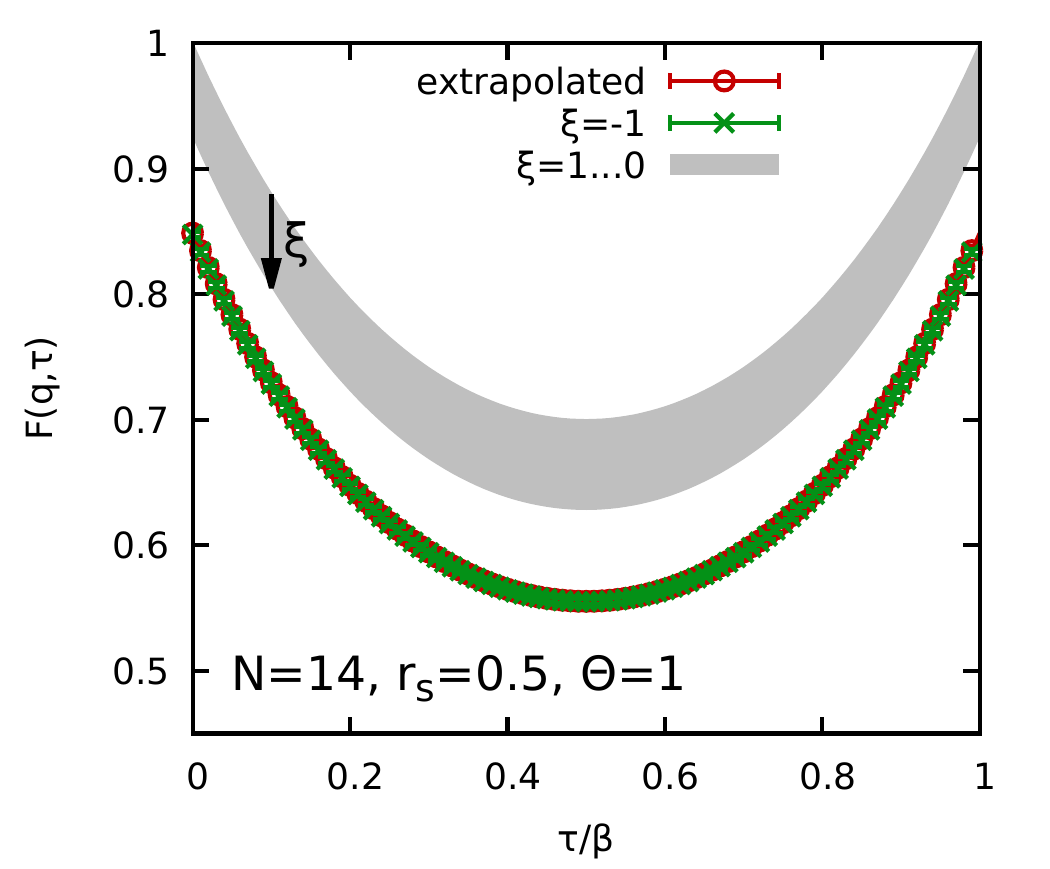}
\caption{\label{fig:UEG_ITCF}
Imaginary-time density correlation function $F(q,\tau)$ of the spin-unpolarized UEG at $r_s=0.5$, $\Theta=1$, $q=1.2q_\textnormal{F}$ and for $N=14$. Shaded grey area: PIMC results for the FSP free domain of $\xi\geq0$; red circles: corresponding extrapolation from the grey area to the fermionic limit of $\xi=-1$ via the quadratic Eq.~(\ref{eq:fit}); green crosses: exact PIMC reference data computed for $\xi=-1$. 
}
\end{figure} 

\begin{figure}
\includegraphics[width=0.462\textwidth]{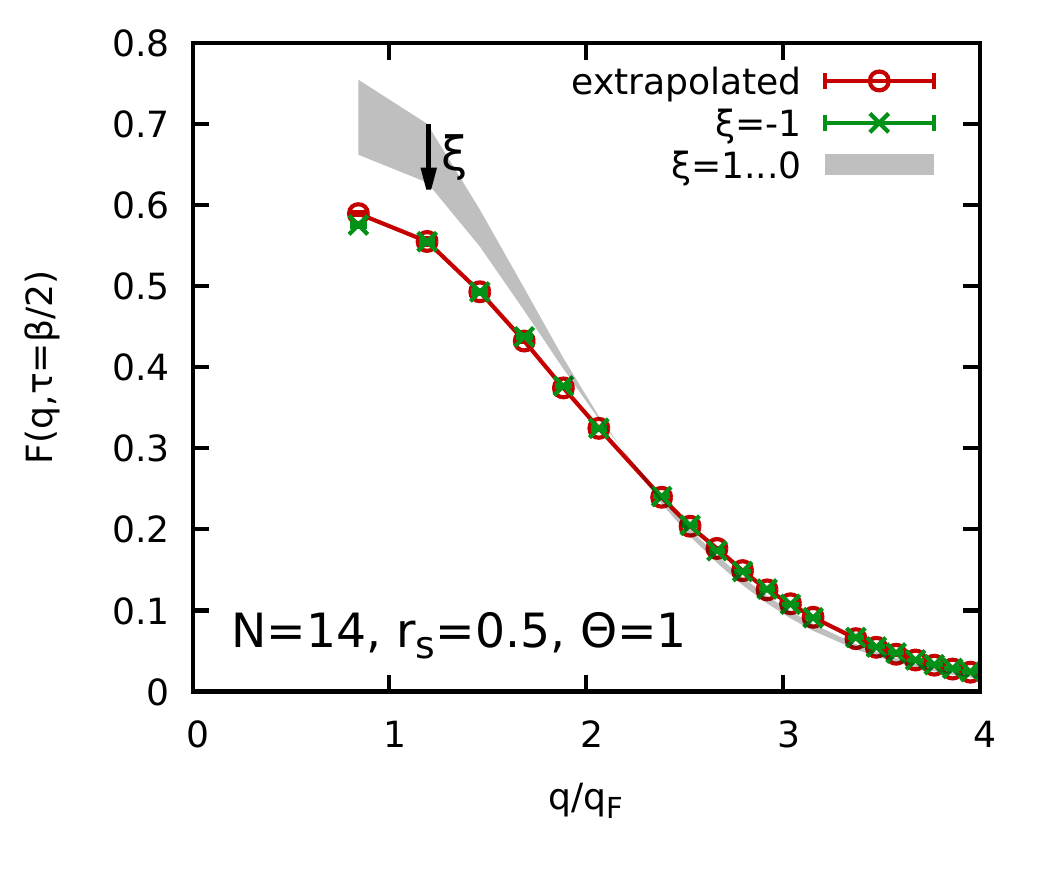}
\caption{\label{fig:UEG_Thermal_SSF}
Thermal static structure factor $S_{\beta/2}(q)=F(q,\beta/2)$ of the spin-unpolarized UEG at $r_s=0.5$, $\Theta=1$ and for $N=14$. Shaded grey area: PIMC results for the FSP free domain of $\xi\geq0$; red circles: corresponding extrapolation from the grey area to the fermionic limit of $\xi=-1$ via Eq.~(\ref{eq:fit}); green crosses: exact PIMC reference data computed for $\xi=-1$. 
}
\end{figure} 

As the culmination of our work, we analyze the extrapolation of the imaginary-time density--density correlation function (ITCF) $F(q,\tau)=\braket{\hat{n}(\mathbf{q},0)\hat{n}(-\mathbf{q},\tau)}$
in Figs.~\ref{fig:UEG_ITCF} and \ref{fig:UEG_Thermal_SSF}; it corresponds to the usual intermediate scattering function $F(q,t)$, but evaluated at an imaginary time argument $t=-i\hbar\tau$, with $\tau\in[0,\beta]$.
From a physical perspective, the ITCF contains exactly the same information as the dynamic structure factor $S({q},\omega)$ to which it is related by a two-sided Laplace transform. Consequently, PIMC results for $F({q},\tau)$~\cite{Boninsegni_maximum_entropy,Saccani_Supersolid_PRL_2012,Filinov_PRA_2016,Dornheim_SciRep_2022,dornheim_dynamic,dynamic_folgepaper} are often used as the basis for an analytic continuation~\cite{JARRELL1996133}, i.e., the numerical inversion of the Laplace transform to infer $\omega$-dependent information from imaginary-time dependent quantum Monte Carlo data.
In addition, it is also possible to directly obtain physical insights from $F(q,\tau)$, see Refs.~\cite{Dornheim_insight_2022,Dornheim_PTR_2022}. This theme has recently been explored by Dornheim \emph{et al.}~\cite{Dornheim_T_2022,Dornheim_T2_2022,Dornheim_review,dornheim2023xray}, who have exploited the symmetry of $F(q,\tau)$ as a highly accurate and model-free temperature diagnostics for X-ray Thomson scattering (XRTS) experiments for WDM.
Finally, we mention the tremendous utility of various ITCFs for the estimation of linear~\cite{dornheim_ML} and nonlinear~\cite{Dornheim_JCP_ITCF_2021} density responses without the need for extensive simulations of a perturbed system.

In Fig.~\ref{fig:UEG_ITCF}, we show the imaginary-time dependence of $F(q,\tau)$ for $q=1.2q_\textnormal{F}$ and $N=14$ with the usual symbol and color coding. It is evident that the $\xi$-extrapolation method is capable to resolve the full $\tau$-dependence of the ITCF with remarkable accuracy. For completeness, we note that the CPIMC method is currently not capable to estimate $F(q,\tau)$, which makes our exact PIMC data for the fermionic limit with $N=14$ the most suitable benchmark.
The analysis of the $\tau$-dependence is complemented by Fig.~\ref{fig:UEG_Thermal_SSF}, where we show results for the thermal static structure factor $S_{\beta/2}(q)=F(q,\beta/2)$ over the entire relevant range of wave numbers.
The extrapolated PIMC results are in excellent agreement with the exact PIMC reference data for $\xi=-1$ everywhere, which nicely confirms the utility of the $\xi$-extrapolation method for the estimation of ITCF properties over the full $\tau$-$q$-plane.

\begin{figure}
\includegraphics[width=0.462\textwidth]{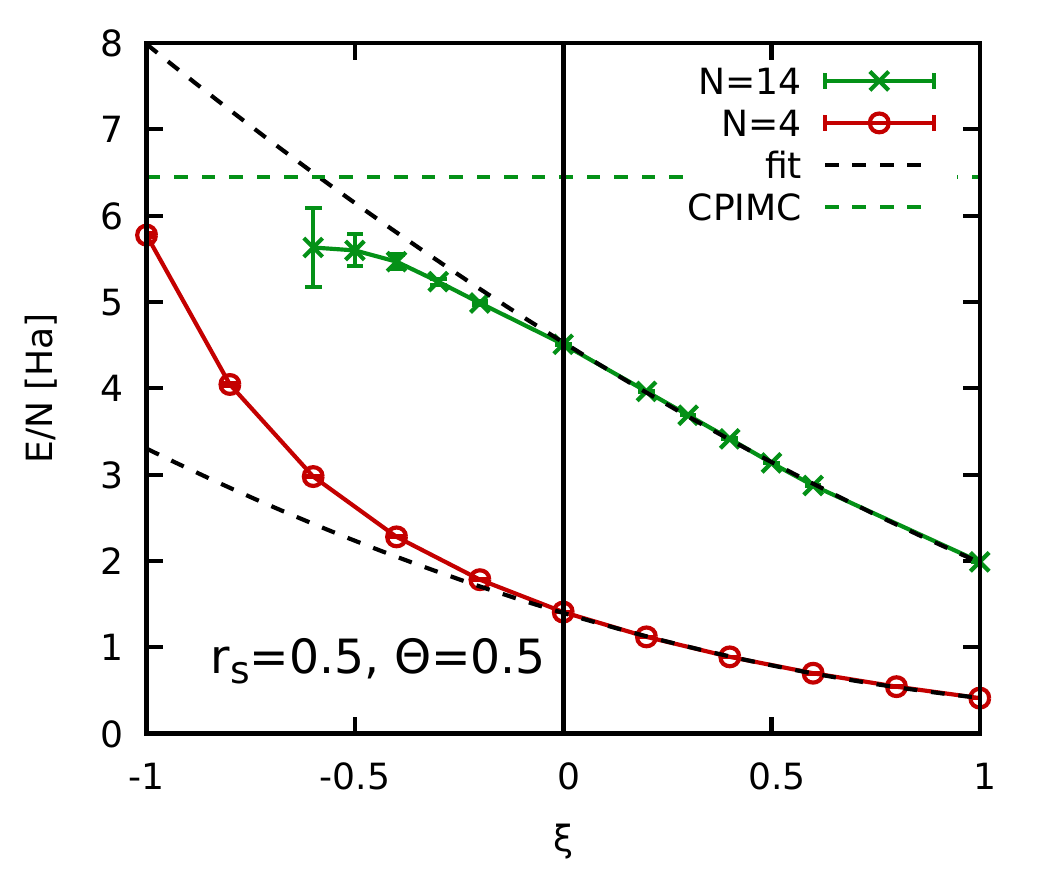}
\caption{\label{fig:UEG_N14_rs0p5_theta0p5}
Energy per particle of the spin-unpolarized UEG at $r_s=0.5$, $\Theta=1$. The red circles and green crosses correspond to PIMC results for $N=4$ and $N=14$, respectively. The dashed black curves show fits based on input data from the FSP free domain ($\xi\geq0$). The horizontal dashed green line indicates the exact CPIMC result at the fermionic limit of $\xi=-1$ for $N=14$.
}
\end{figure}

To conclude this analysis of the extrapolation method applied to the UEG, we consider the more strongly quantum degenerate case of $r_s=0.5$ and $\Theta=0.5$ in Fig.~\ref{fig:UEG_N14_rs0p5_theta0p5}. Specifically, the red circles have been obtained for $N=4$ and exhibit a strongly non-linear dependence on the extrapolation parameter $\xi$. Consequently, the corresponding extrapolation from $\xi\geq0$ via the quadratic Eq.~(\ref{eq:fit}), i.e., the dashed black line, becomes inaccurate even for $\xi\gtrsim0.2$, and underestimates the true fermionic limit of the energy by nearly $50\%$. For $N=14$, PIMC simulations are only feasible for $\xi\lesssim0.4$ due to the FSP. Still, the corresponding extrapolation starts to deviate from the green crosses for $\xi>0$, and the thus predicted fermionic limit of $\xi=-1$ substantially disagrees from the exact CPIMC reference value (the horizontal dashed green line). In fact, the observed behaviour strongly suggests the presence of an inflection point in the $E(\xi)$ curve for $N=14$ around $\xi=0.6$, which, however, cannot be possibly reproduced by the quadratic polynomial form of Eq.~(\ref{eq:fit}).

\section{Summary and Discussion\label{sec:summary}}

In this work, we have investigated in detail the utilization of \emph{ab initio} PIMC simulations with a continuous interpolating variable $\xi\in[-1,1]$ as a tool for the description of weakly quantum degenerate Fermi systems without the exponential computational bottleneck as a result of the usual sign problem. This has been achieved on the basis of extensive new simulation results for two representative model systems: i) electrons in a 2D harmonic oscillator potential, which constitute a finite and strongly inhomogeneous system, ii) the warm dense uniform electron gas, which serves as the archetypal electronic bulk system. Overall, our findings are qualitatively very similar for both model systems. The extrapolation from the sign-problem free domain ($\xi\geq0$) to the fermionic limit of $\xi=-1$ proposed by Xiong and Xiong~\cite{Xiong_JCP_2022} works very well for weakly quantum degenerate systems. In this regime, it is possible to accurately infer the properties of the given Fermi system of interest without the exponential increase in computation time due to the usual fermion sign problem. In particular, we have demonstrated that the method is capable to give highly accurate results ($\sim0.1\%$) for the UEG at $r_s=0.5$ and $\Theta=1$ even for $N=66$ electrons, resulting in a speed-up exceeding eleven orders of magnitude; these conditions are otherwise fully inaccessible to the direct PIMC method, with an average sign of $S\sim10^{-6}$.

In addition, we have extended the thermodynamic considerations of the original Ref.~\cite{Xiong_JCP_2022} by presenting the first extrapolation-based results for observables beyond the total energy, namely the radial density profile $n(r)$ for harmonically confined electrons and the static structure factor $S(q)$ as well as the imaginary time density--density correlation function $F(q,\tau)$ for the bulk UEG. In the harmonic trap, we found that the extrapolation success can depend on the local degree of quantum degeneracy, which is interesting in its own right. In the UEG, the extrapolation of $S(q)$ [$F(q,\tau)$] has worked exceptionally well for the entire relevant range of wave numbers [over the entire $\tau$-$q$-plane], which makes the extrapolation scheme a very flexible tool for the investigation of large, weakly degenerate warm-dense matter systems that are inaccessible to the direct PIMC method due to the sign problem.

Despite these remarkable successes, it is important to note that the extrapolation method systematically breaks down when the degree of quantum degeneracy becomes substantial. In particular, we find that the functional dependence of observables such as the total energy on the interpolating parameter $\xi$ exhibits an inflection point in the strongly degenerate limit, which cannot be reproduced by the quadratic fit function of Eq~(\ref{eq:fit}). Unfortunately, the method breaks down even in moderately degenerate cases without such an inflection point. In practice, we find that the best criterion to decide if the extrapolation method will be reliable is to carry out PIMC simulations of a small system at the same conditions over the entire $\xi$-range. This strategy seems to be particularly reliable for bulk systems such as the UEG, where the overall degree of quantum degeneracy will only weakly depend on the particular number of particles.

We are convinced that our in-depth analysis of the extrapolation method originally proposed by Xiong and Xiong~\cite{Xiong_JCP_2022} opens up a range of avenues for future investigations. First, the successful performance of the extrapolation approach for the weakly degenerate UEG makes it a  promising candidate for the investigation of highly ionized warm-dense matter systems such as hydrogen, helium, and potentially even heavier elements. We note that the corresponding extreme densities and temperatures are particularly relevant for upcoming experiments with inertial confinement fusion as they are realized for example at the National Ignition Facility~\cite{Moses_NIF,Tilo_Nature_2023,MacDonald_POP_2023,boehme2023evidence}.
In addition, the capability of the approach to give accurate results for large Fermi systems in the regime of weak degeneracy without the exponential increase in the compute time due to the sign problem makes it useful to test existing schemes for finite-size corrections, and to develop new approaches for observables for which currently no finite-size corrections exist. Moreover, large-scale PIMC simulations of WDM will be useful as a benchmark for computationally less demanding though approximate density functional theory simulations, which currently constitute the work horse in this field~\cite{wdm_book,Dornheim_review}.

An additional interesting avenue for future research is the application of the method to a variety of different observables. A particularly interesting class of properties is given by different imaginary-time correlation functions, which give access to a host of linear and non-linear response properties of a given quantum system~\cite{Dornheim_JCP_ITCF_2021,Dornheim_review}, and, additionally, serve as key input for the analytic continuation~\cite{JARRELL1996133} to dynamic properties such as the dynamic structure factor $S(\mathbf{q},\omega)$. Moreover, the method is in no way restricted to Coulomb interacting electrons that have been the focus of the present study. For instance, it is known that ultracold $^3$He only weakly depends on the specific type of quantum statistics for temperatures as low as $T=2\,$K~\cite{Dornheim_SciRep_2022}, making the application of the extrapolation scheme promising in this regime.

Furthermore, we point out that the development of the extrapolation technique itself remains in its infancy. While some higher-order polynomials were tested by Xiong and Xiong~\cite{Xiong_JCP_2022}, improved extrapolation ans\"atze should be investigated. Future methodological improvements might include the introduction of better fit functions that are capable to cover and, hopefully, even predict the inflection point in the quantum degenerate regime, and the combination of the extrapolation with other tools for the PIMC/PIMD simulation of fermions such as the permutation blocking paradigm~\cite{Dornheim_NJP_2015}.

Finally, let us discuss the enticing possibility of alternative fictitious PIMC setups, whose empirical extrapolation procedure can even be guided by established theoretical approaches. The present fictitious PIMC technique can be perceived as simulation-based, since it introduces an artificial penalty on pair-particle exchanges directly on the canonical partition function, which appropriately interpolates between the well-known fermionic and bosonic limits. Physics-based fictitious PIMC techniques could be developed that are based on fractional exclusion statistics. It is reminded that fractional (often also coined as anyonic) statistics naturally arise in two dimensions~\cite{Wilczek_PRL_1982,Wu_PRL_1984,Forte_1992}. To be more specific, in the prevailing theory of the fractional quantum Hall effect, the fractionally charged excitations of the Laughlin states are anyons, since the phase shift arising in their many-body wavefunction owing to the exchange of any two identical quasiparticles can be assigned any value~\cite{quantum_theory,Wilczek_book}. This motivated general formulations of quantum statistical mechanics for three-dimensional non-interacting gases with occupation number distributions that interpolate between bosons and fermions~\cite{Avinash_book}. The fractional elements are initially inserted in the combinatorial expression for the counting of many-body states, whose constrained optimization yields generalized equilibrium distributions that naturally interpolate between the Fermi-Dirac and the Bose-Einstein distributions~\cite{Haldane_PRL_1991,Wu_PRL_1994,Polychronakos_1996}. Such fractional equilibrium distributions can be straightforwardly incorporated in different schemes of the self-consistent dielectric formalism known to be accurate in the warm dense matter~\cite{stls_original,vs_original,stls,schweng,stls2} and the strongly coupled regimes~\cite{tanaka_hnc,castello2021classical,Tolias_JCP_2021,Tolias_JCP_2023}. Fractional equilibrium distributions can also be incorporated in the classical mapping formalism, where they will affect the Fermi-hole potential that is introduced from the non-interacting limit aiming to treat quantum exclusion effects~\cite{perrot,Dharmawardana,Dutta_2013}. The interaction energy or static structure factor output from these fractionally extended formalisms can then be utilized for the construction of empirical functional forms that connect the bosonic and fermionic sectors. Naturally, the corresponding canonical partition function would need to employed in the respective fictituous PIMC technique.

\section*{Acknowledgments}
This work was partially supported by the Center for Advanced Systems Understanding (CASUS) which is financed by Germany’s Federal Ministry of Education and Research (BMBF) and by the Saxon state government out of the State budget approved by the Saxon State Parliament. 
This work has received funding from the European Research Council (ERC) under the European Union’s Horizon 2022 research and innovation programme
(Grant agreement No. 101076233, "PREXTREME").
The PIMC calculations were partly carried out at the Norddeutscher Verbund f\"ur Hoch- und H\"ochstleistungsrechnen (HLRN) under grant shp00026, and on a Bull Cluster at the Center for Information Services and High Performance Computing (ZIH) at Technische Universit\"at Dresden.

\bibliography{bibliography}
\end{document}